\documentclass[a4paper,10pt]{article}
\usepackage{jheppub,lipsum,ulem,hyphenat,xcolor,tcolorbox,slashed,float,bbold,bm,comment,enumitem,multicol,multirow,mathtools,sidecap,tikz,booktabs,changepage,graphicx,subcaption}
\usepackage[T1]{fontenc}
\usepackage[compat=1.1.0]{tikz-feynman}
\graphicspath{ {figures/} }
\pdfoutput=1  
\allowdisplaybreaks

\let\emph\textit

\definecolor{lime}{HTML}{A6CE39}
\DeclareRobustCommand{\orcidicon}{\hspace{-2.1mm}
\begin{tikzpicture}
\draw[lime, fill=lime] (0,0) circle [radius=0.13] node[white] {{\fontfamily{qag}\selectfont \tiny \,ID}}; \draw[white, fill=white] (-0.0525,0.095) circle [radius=0.007]; 
\end{tikzpicture} \hspace{-3.2mm} }
\foreach \x in {A, ..., Z}{\expandafter\xdef\csname orcid\x\endcsname{\noexpand\href{https://orcid.org/\csname orcidauthor\x\endcsname} {\noexpand\orcidicon}}}

\title{Unveiling the Secrets of New Physics Through Top Quark Tagging}
\author[a,b]{Rameswar Sahu,\orcidC{}}
\author[c,d]{Saiyad Ashanujjaman,\orcidA{}}
\author[a,b]{Kirtiman Ghosh,\orcidB{}}

\affiliation[a]{Institute of Physics, Bhubaneswar, Sachivalaya Marg, Sainik School Post, Bhubaneswar 751005, India}
\affiliation[b]{Homi Bhabha National Institute, Training School Complex, Anushakti Nagar, Mumbai 400094, India}
\affiliation[c]{Institute of High Energy Physics, Chinese Academy of Sciences, Beijing 100049, China}
\affiliation[d]{Kaiping Neutrino Research Center, Jiangmen 529399, China}
\emailAdd{rameswar.s@iopb.res.in}
\emailAdd{kirti.gh@gmail.com}
\emailAdd{s.ashanujjaman@gmail.com}
\abstract{The ubiquity of top-rich final states in the context of beyond the Standard Model (BSM) searches has led to their status as extensively studied signatures at the LHC. Over the past decade, numerous endeavours have been undertaken in the literature to develop methods for efficiently distinguishing boosted top quark jets from QCD jets. Although cut-based strategies for boosted top tagging, which rely on substructure information from fat jets resulting from the hadronic decay of boosted top quarks, were introduced in the literature as early as 2008, recent years have witnessed a surge in the utilization of machine learning-based approaches for the classification of top-jets from QCD jets. The review focuses on the present status of boosted top tagging and its application for BSM searchers.}


\begin{document} 

\maketitle
\flushbottom
\section{Introduction}
The Large Hadron Collider (LHC) \cite{Evans:2008zzb} at CERN represents the pinnacle of scientific engineering, dedicated to unraveling the fundamental constituents of nature. This proton-proton collider, designed to probe the tiniest structures within a controlled laboratory setting, delivers collisions at unparalleled center-of-mass energies: 7 and 8 TeV for Run I, 13 TeV for Run II, and 14 TeV from Run III onwards. Although protons are not elementary particles themselves, they are composed of quarks and gluons. Quantum Chromodynamics (QCD), the theory governing strong interactions within the Standard Model (SM) of particle physics, describes the interaction of quarks and gluons. Despite the challenge of directly detecting free quarks and gluons due to color confinement, they remain pivotal in discussions regarding high-energy hadron collider phenomenology. When quarks or gluons are produced at high energies, they promptly fragment and hadronize, resulting in a collimated spray of energetic particles known as a jet. By measuring the energy and direction of these particles (the jet) one can glean insights into the properties of the original parton. Defining a jet involves a prescription (jet algorithms) \cite{Sterman:1977wj,JADE:1986kta,JADE:1988xlj,Catani:1991hj,Catani:1993hr,Ellis:1993tq,Dokshitzer:1997in,Cacciari:2008gp} to group hadrons into jets and assign momentum to the resulting jet. In addition to the collimated beam of hadrons resulting from the hadronization of light quarks and gluons (light jets), the hadronic decay of boosted heavy SM particles such as the $W/Z$-boson, Higgs boson ($h$), or top quark also leads to a collimated spray of hadrons that can resemble a single jet. 

Since its inception, the LHC has pursued evidence of physics beyond the SM (BSM). Despite the remarkable discovery of the Higgs boson \cite{ATLAS:2012yve,CMS:2012qbp}, which validated aspects of the SM, the absence of concrete evidence supporting BSM physics has prompted researchers to venture into higher energy regimes. These enhanced energies enable the production of boosted heavy SM particles like the top quark, $W/Z$-boson, and Higgs boson. The hadronic decays of these boosted SM particles result in a collimated cluster of quarks, forming large-radius (large-R) jets known as ``fat jets'' with distinctive sub-structure features. 
There are several advantages to designing search strategies for heavy BSM resonances that decay into highly Lorentz-boosted massive SM particles when hadronic decays of these boosted particles are considered. For one, the hadronic decays of these particles result in a higher signal rate compared to their leptonic decays. Further, the visibility of the hadronic decay products of these particles at the LHC detectors enables the kinematic reconstruction of the decay cascade. More importantly, only a small fraction of the SM background events would give rise to fat jets in the final state. Consequently, fat jet final state signatures are beset with considerably less SM background than those with resolved jets. As a result, the analysis of jet substructure at the LHC resulting from the hadronic decay of boosted top quarks, $W/Z$-bosons, or the Higgs boson has been instrumental in searching for heavy BSM resonances across various new physics scenarios, including supersymmetry \cite{ATLAS:2021yqv, ATLAS:2020xgt, Ghosh:2012ud, ATLAS:2020dsf, ATLAS:2022ihe}, extra-dimensional models \cite{CMS:2018rkg, ATLAS:2019npw}, leptoquark models \cite{ATLAS:2019qpq, CMS:2020wzx}, and other extensions of the SM \cite{ATLAS:2023ibb, ATLAS:2023qqf, ATLAS:2023taw, Ashanujjaman:2022cso, Ashanujjaman:2021zrh, CMS:2019eqb, ATLAS:2022ozf, CMS:2018dcw, ATLAS:2019npw, ATLAS:2020lks}. Efficiently identifying the particle origin of fat jets is crucial to enhancing the sensitivity of the LHC and future colliders. This necessitates a significant shift in analysis strategy and the development of new innovative methodologies for tagging the particle origin of the fat-jets.

In this review, our focus is on classifying fat jets resulting from the hadronic decay of boosted top quarks and distinguishing them from light quarks and gluon jets (hereafter referred to as QCD jets). Top quarks at the LHC are particularly intriguing due to the substantial $t\bar t$ production cross-section, essentially making the LHC a ''top factory''. The millions of top quarks produced at the LHC are expected to provide insights into the SM and its potential extensions. While most top quarks are produced near threshold and can be identified using traditional top reconstruction algorithms similar to those used at the Tevatron, some top quarks produced at the LHC are highly boosted. Theoretical interest in top quarks is heightened by their large Yukawa coupling. The large top Yukawa coupling not only plays a critical role in computing electroweak precision observables \cite{Baak:2014ora} and determining the vacuum stability \cite{Degrassi:2012ry} of the Standard Model (SM), but it also has a notable impact on the masses and interactions of various BSM resonances. Many of these resonances exhibit enhanced couplings with the top quark, contributing to a final state rich in top quarks at the LHC. Over the past decade, considerable efforts have been made in the literature to develop effective methods for efficiently distinguishing boosted top quark jets from QCD jets. While early literature introduced cut-based strategies for boosted top tagging \cite{Thaler:2008ju, Kaplan:2008ie}, leveraging substructure information from fat jets resulting from the hadronic decay of boosted top quarks, recent years have seen a surge in the adoption of machine learning-based approaches for top-jet classification. In this article, we provide a comprehensive review of various cut-based and machine learning-based approaches proposed in the literature over the past couple of decades for top-tagging. This review aims to synthesize the advancements in top-jet classification methodologies, highlighting their evolution and effectiveness in distinguishing top quark jets from QCD jets at high-energy collider experiments like the LHC.

The review is organized as follows: In the next section, we review high-level feature (HLF) based top classifiers. Sections 3 and 4 are dedicated to Image-based Classifiers and Graph Neural Network (GNN) classifiers, respectively. In section 5, we provide a list of BSM scenarios that result in boosted top quark final states at the LHC. Finally, we summarize our findings in section 6.

\section{High-Level Feature (HLF) based classifiers}
\label{sec:HLF}
Tagging boosted objects is a long-pursued quest dating back to the eras of the Tevatron.  These boosted objects (or, in the context of our discussion, boosted fat jets) can have different origins ranging from decays of SM bosons (W/Z/H), the top quark, light quarks/gluons, or some BSM particles. As discussed in the introduction, we will focus our attention on the tagging of boosted top jets, i.e., identifying fat jets originating from hadronic decay of top quarks from those originating from QCD-initiated light quarks and gluon jets. Initial works in this direction rely on identifying b-jets inside the top jet and the reconstruction of the invariant mass of the W-boson inside the top and top quark mass as a whole. The problem with this approach was the isolation of the b-jet and light jets, which are highly collimated due to the boosted mother particle. This makes the procedure inefficient in scenarios where the production cross-section of these fat jets is small. This led to an in-depth investigation of the jet sub-structure and resulted in the development of several jet substructure variables/high-level features (HLFs).  \\

The literature on jet substructure variables is vast. Some interesting examples include the jet energy moments \cite{Gur-Ari:2011cjr}, the energy correlation functions (ECFs) \cite{Larkoski:2013eya}, the generalized energy correlation functions (ECFGs) \cite{Moult:2016cvt}, N -subjettiness variables \cite{Thaler:2010tr, Thaler:2011gf, Stewart:2010tn}, and Energy Flow Polynomials \cite{Komiske:2017aww}. For a comprehensive discussion, we direct the interested readers to reference \cite{Kogler:2018hem,Abdesselam:2010pt,Altheimer:2012mn,Altheimer:2013yza,Adams:2015hiv,Marzani:2019hun,Larkoski:2017jix}. We have hand-picked some of these and some physics-inspired algorithmic approaches for the subsequent discussion.\\

\begin{figure}[!htb]
	\centering
	\includegraphics[width=0.8\columnwidth]{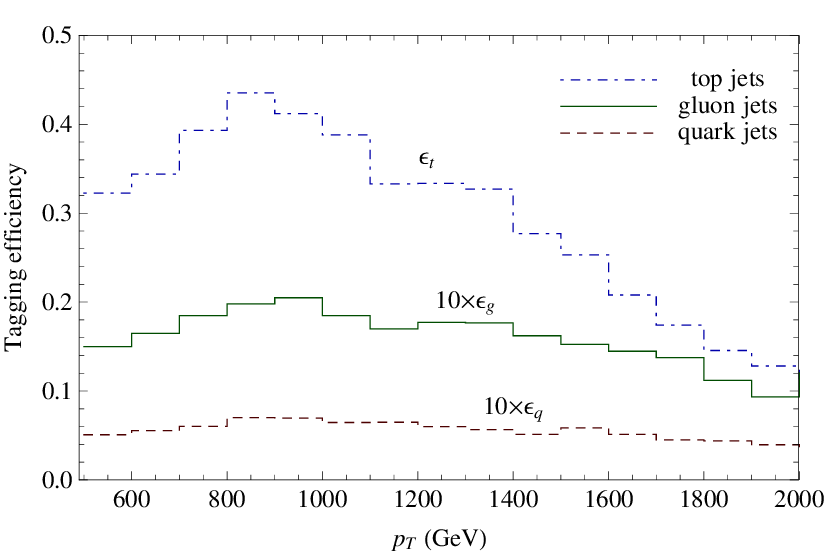}
	\caption{ The variation of top tagging efficiency and the mis-tagging efficiencies of gluon and light-quark jets with the transverse momentum of the fat jets \cite{Kaplan:2008ie}.}
	\label{fig:JHTT}
\end{figure}

The Johns Hopkins Top Tagger (JHTT), originally introduced in reference \cite{Kaplan:2008ie}, looks into the subjet structure of the fat jet, applying additional kinematic criteria to identify fat jets originating from top quarks. To begin with, C/A fat jets with a given R parameter are considered. These fat jets undergo a sequential declustering procedure. First, the fat jet J is declustered into two subjets, $j_1$ and $j_2$. If the softer subjet is discarded if it has $p_{T,j} < \delta_p \times p_{T, J}$, for some predefined $\delta_p$ while the harder subjet undergoes further declustering until certain conditions are met (if both subjets do not fulfill above criteria they both are considered for further analysis). These conditions include the subjet being comprised of a single calorimeter cell, the daughter jets from declustering being too close ($|\eta|+|\phi| < \delta_r$), or both daughter jets satisfying the initial criteria of $p_{T,j} < \delta_p \times p_{T, J}$.  Fat jets with 3 or 4 subjets are considered for further analysis. For their final analysis, reference \cite{Kaplan:2008ie} uses SM top pair production and QCD di-jet production for generating signal and background fat jets in Pythia \cite{Bierlich:2022pfr}. To incorporate detector effects, final state visible particles are combined in grids of size $0.1 \times 0.1$ and passed onto the clustering algorithm. Only fat jets with $p_T > 500$ GeV and $|\eta| < 2.5$ are retained. Different R-parameter values are adopted depending on the event's scalar $E_T$. For $E_T >$ 1000, 1600, 2600 GeV they select R=0.8, 0.6, 0.4, $\delta_p = 0.1, 0.05, 0.05$, and $\delta_r = 0.19, 0.19, 0.19$, respectively. Moreover, the fat jets must also satisfy $p_{T, J} > 0.7 \times E_T/2$. Once the final subjets are identified, they pass through some kinematic requirements. For jets with $p_T < 1000$ GeV, these amounts to the requirement that the invariant mass of the final subjets must be within 145-205 GeV, and there must be two subjets with invariant mass in the window of 65 to 95 GeV. For fat jets with $p_T > 1000$ GeV, the upper window is shifted to $p_T/20$ + 155 GeV and $p_T/40 + 70$ GeV respectively for top and W mass reconstruction. Additionally, the reconstructed W helicity angle must adhere to $cos\theta_h <$ 0.7 in both scenarios. For a detailed discussion, we encourage the interested reader to refer to the original paper \cite{Kaplan:2008ie}. For completeness, we present their final results in Figure \ref{fig:JHTT}. \\

\begin{figure}[!htb]
	\centering
	\includegraphics[width=0.3\columnwidth]{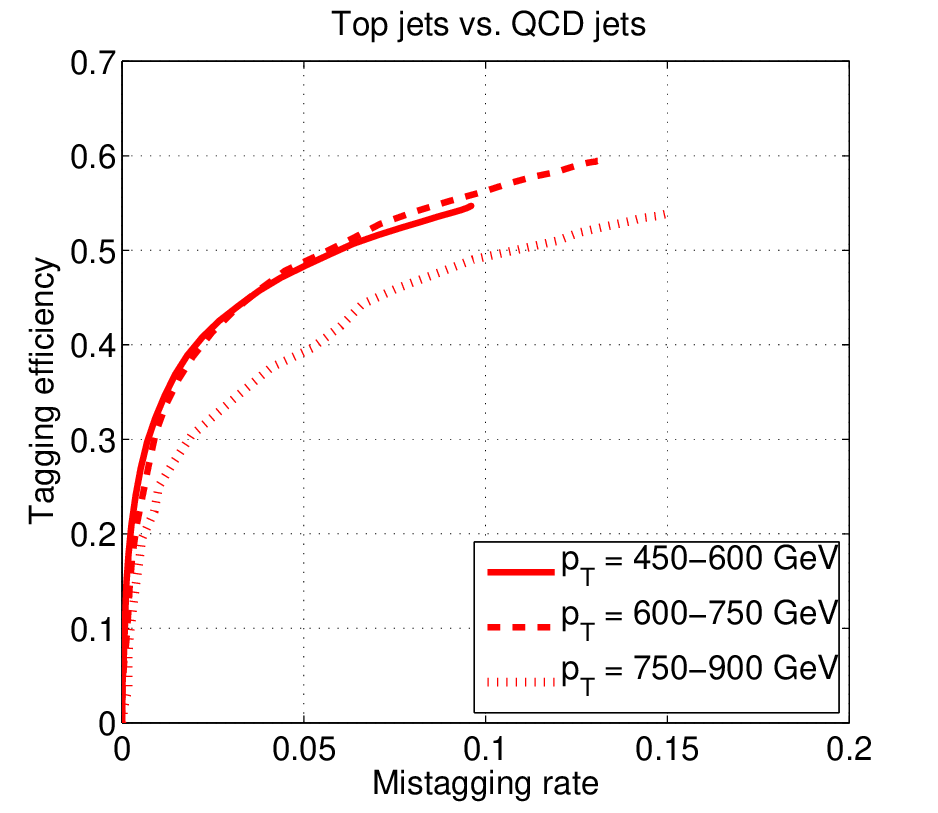}
	\includegraphics[width=0.3\columnwidth]{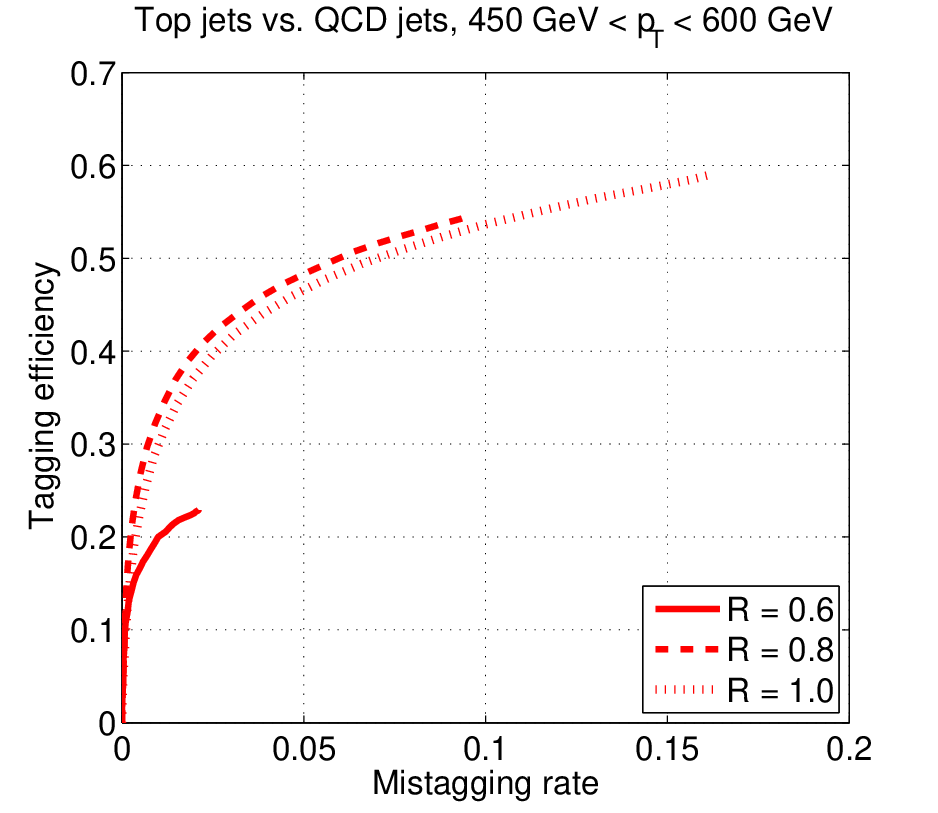}
	\includegraphics[width=0.3\columnwidth]{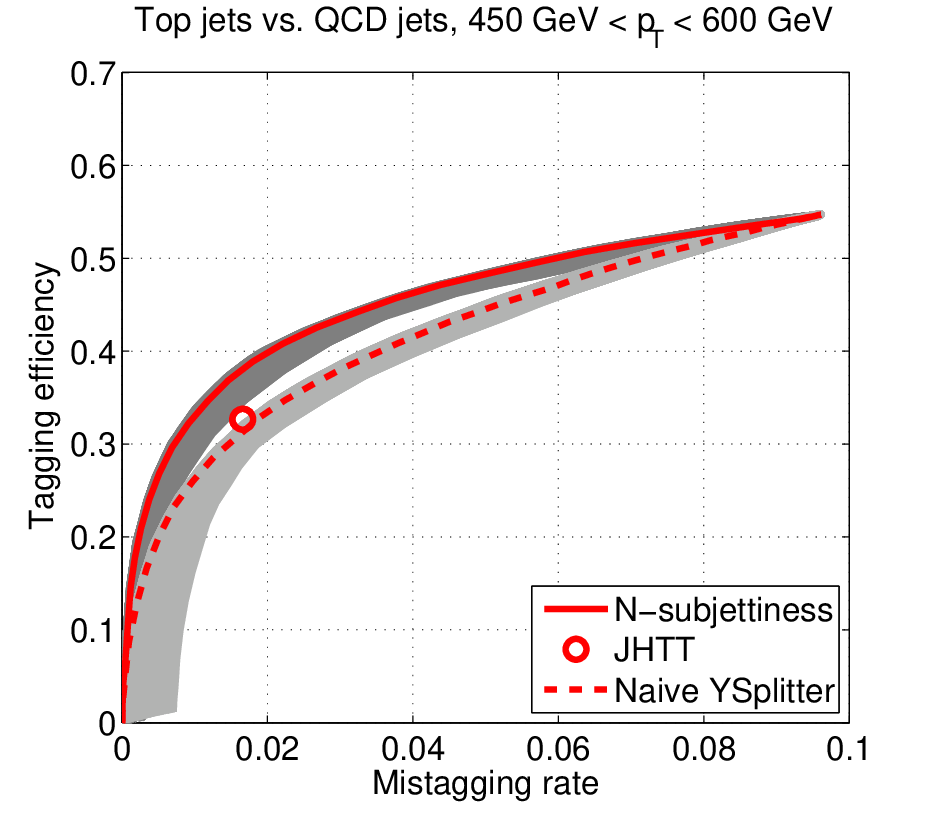}
	\caption{ The variation of top tagging efficiency with the mis-tagging rate of gluon and light-quark jets. The plot on the left shows the variation of classification performance with $p_T$ of the fat jet \cite{Thaler:2010tr}. The plot in the middle shows the dependence on the radius of reconstruction for fat jets in the transverse momentum range of 450-600 GeV. The rightmost plot compares the performance of the N-subjettiness tagger with that of the Johns Hopkins Top Tagger (JHTT) \cite{Kaplan:2008ie} and the YSplitter method \cite{Butterworth:2002tt, Brooijmans:2008zza}.}
	\label{fig:Nsubjet}
\end{figure}

N-subjettiness is an inclusive jet shape originally introduced in reference \cite{Thaler:2010tr}. It is designed to identify the energy deposition pattern inside a fat jet and quantify the major sources of these energies. In simpler terms, it quantifies the number of prongs of a fat jet. To define N-subjettiness, one needs information on the N candidate subjets inside a fat jet. To remain IRC safe, reference \cite{Thaler:2010tr} employs the exclusive-$k_{T}$ algorithm for this task. Once these subjets are known, the N-subjettiness can be calculated via the relation \cite{Thaler:2010tr}:
\begin{equation}
\tau_N = \frac{\sum_k p_{T,k}~ \text{min}\{\Delta R_{1,k},...\Delta R_{N,k}\}}{\sum_k p_{T,k} R_0}
\end{equation}
Here, $R_0$ is the R-parameter of the fat jet clustering algorithm. The sum runs over all the jet constituents of momentum $p_{T,k}$ and $\Delta R_{N,k}$ represents the separation in the rapidity-azimuth plane between the $N$th subjet and the $k$th constituent. In an ideal setting, $\tau_N \approx 0$ when all the constituents are aligned with some candidate subjet. As argued in reference \cite{Thaler:2010tr}, the ratio of N-subjettiness variables is a better candidate for fat-jet tagging and will be used in the subsequent discussion. To test the effectiveness of the $\tau_N$ variable, reference \cite{Thaler:2010tr} has demonstrated its application in separating top jets from QCD background jets. After generating top-pair events in Pythia-8.135 \cite{Sjostrand:2007gs}, a simple detector simulation is performed where visible final state particles with $\eta < 4$ are combined to form $0.1\times 0.1$ sized calorimeter cells. These cells are assumed massless and are used in Fastjet 2.4.2 \cite{Cacciari:2005hq} for reconstructing the fat jets. Fat jets in different ranges of transverse momenta (with $\eta < 1.3$) clustered using different values of $R_0$ are considered for the final analysis. To effectively suppress the QCD background, an additional criterion of 145 GeV $< m_{jet} <$ 205 GeV is imposed on the fat jet mass. For a detailed discussion on the signal and background event generation, we direct the interested reader to reference \cite{Thaler:2010tr}. For our discussion, we present their results in figure \ref{fig:Nsubjet}. \\

\begin{figure}[!htb]
	\centering
	\includegraphics[width=0.4\columnwidth]{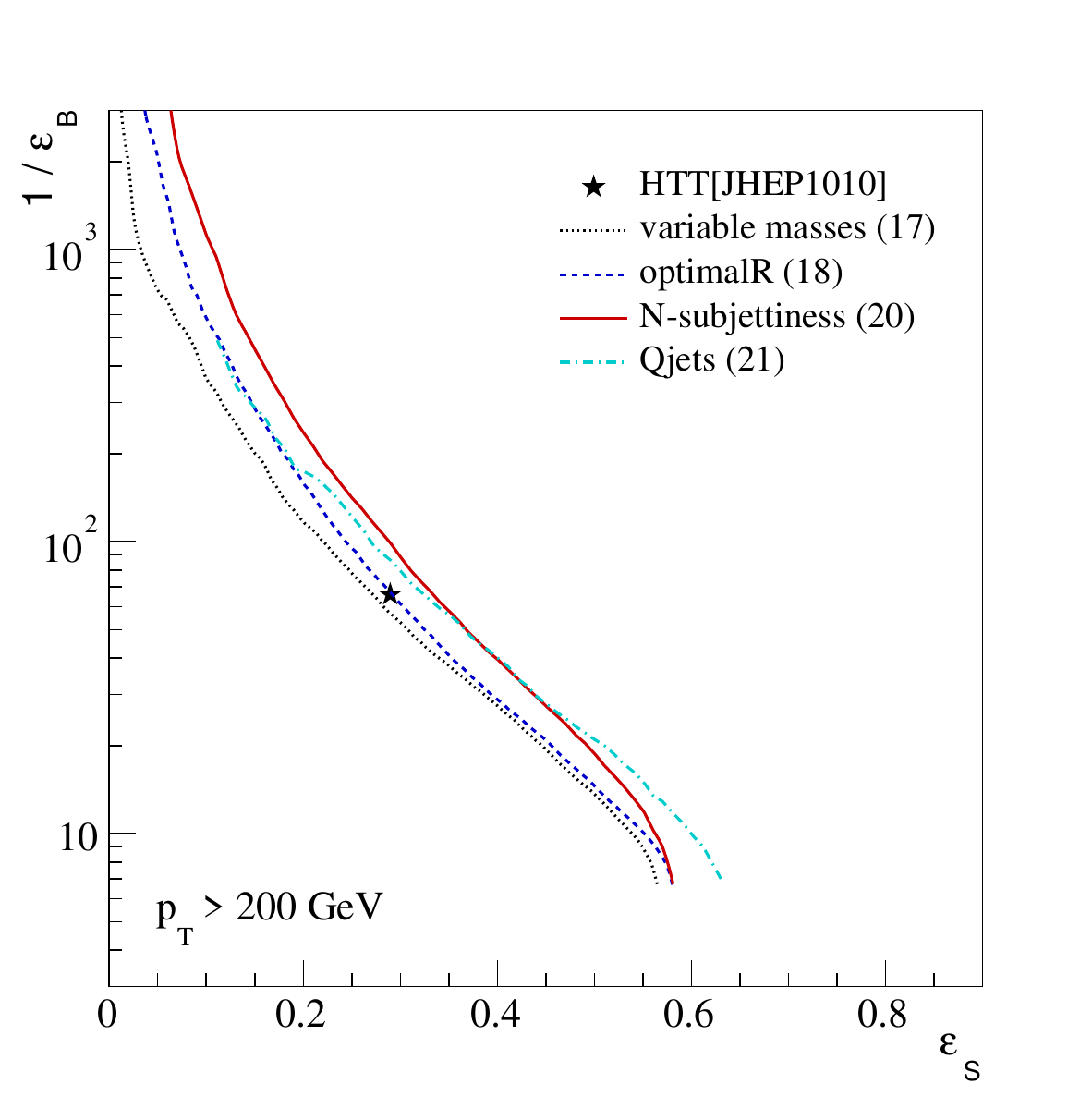}
	\includegraphics[width=0.4\columnwidth]{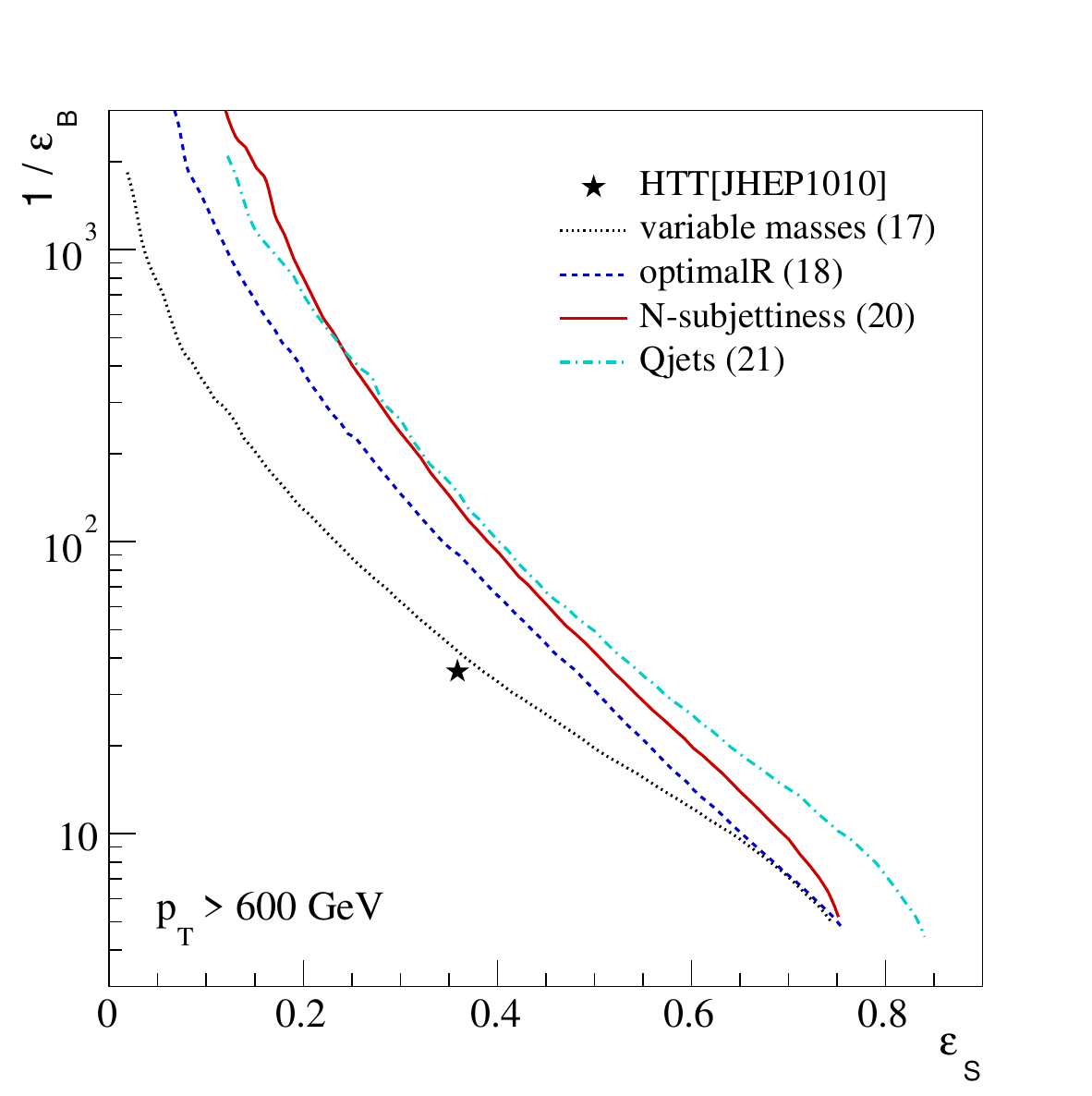}
	\caption{The ROC curves for the four BDT classifiers \cite{Kasieczka:2015jma}. For comparison, the performance of the previous version of HTT \cite{Plehn:2010st} is shown as a star pointer. The plot on the left is for fat jets with $p_T > 200$ GeV while that on the right is for $p_T > 600$ GeV.}
	\label{fig:HTT}
\end{figure}

The HEPTopTagger (HTT), initially introduced in reference \cite{Plehn:2009rk}, has undergone several modifications in subsequent works \cite{Plehn:2010st, Kasieczka:2015jma}. For our discussion, we adopt the version outlined in reference \cite{Kasieczka:2015jma}, known as HEPTopTagger2 (HTT2). While HTT2 shares some steps with the Johns Hopkins Top Tagger, like using C/A jets and gradual declustering to identify relevant subjets, it incorporates several physics-motivated procedures to enhance top tagging performance. Unlike the JHTT, HTT2 addresses both low ($p_T >$ 200 GeV) and high ($p_T >$ 600 GeV) $p_T$ fat jets. A complete description of the algorithm is beyond the scope of our discussion, and we direct the interested reader to ref \cite{Kasieczka:2015jma} for a comprehensive discussion. Here, we only highlight its important features. The algorithm starts with a large radius (R=1.8) C/A fat jet and implements a mass drop criterion to identify the relevant subjets. It then enforces a resolution cut and selects the five hardest subjets, which are further clustered into three candidate subjets meeting specific mass criteria $m_{rec} \in [150,200]$ GeV. To qualify as top constituents, the paired invariant masses of these three subjets are required to satisfy some additional conditions \cite{Kasieczka:2015jma}. If multiple triplets satisfy these criteria, the one with $m_{rec}$ closest to the top pole mass is kept for further analysis. The transverse momentum of the reconstructed top jets must satisfy $p_T >$ 200 GeV. To optimize performance, four Boosted Decision Trees (BDTs) are trained, each focusing on distinct input features, and accordingly, they are named variable masses, OptimalR, N-subjettiness, and Qjets. For more details, see ref \cite{Kasieczka:2015jma}. We present their classification performance for $p_T > 200$ GeV and $p_T > 600$ GeV fat jets in figure \ref{fig:HTT}. \\

\begin{figure}[!htb]
	\centering
	\includegraphics[width=0.8\columnwidth,width=0.4\textheight]{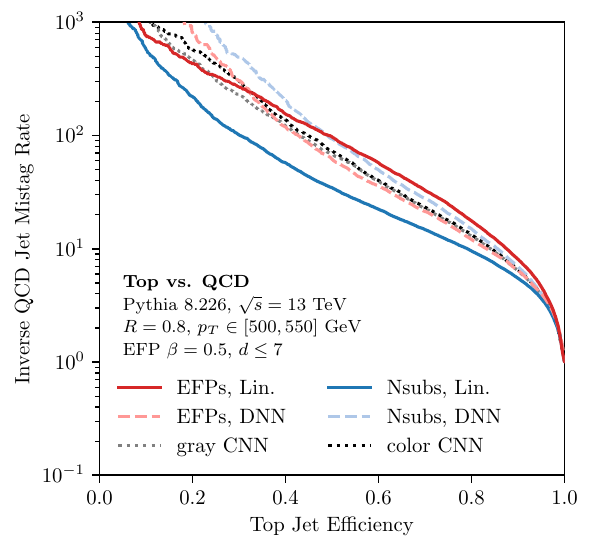}
	\caption{The ROC curves for the EFP-based (red) liner (solid) and DNN (dotted) classifiers and the ROC curves of N-subjettiness (blue) and image-based (gray) classifiers \cite{Komiske:2017aww}.}
	\label{fig:PFP}
\end{figure}

Energy Flow Polynomials (EFPs) for jet tagging were originally introduced in reference \cite{Komiske:2017aww}. They form a liner basis for IRC-safe observables, making them suitable for use with linear classifiers. As demonstrated in ref \cite{Komiske:2017aww}, EFPs have a one-to-one correspondence with loopless multigraphs. This characteristic makes constructing the basis of EFPs much simpler. It also allows the truncation of the infinite number of EFPs at any particular order determined by the number of vertices in the multigraph. In terms of a multigraph G with N vertices, the EFPs take the form \cite{Komiske:2017aww}:
\begin{equation}
EFP_G = \sum_{i_1=1}^M...\sum_{i_N=1}^M z_{i_1}...z_{i_N} \prod_{(k,l)\in G} \theta_{i_k i_l}
\end{equation}
where $(k, l)\in G$ are the edges of the multigraph. M denotes the number of particles in the jet, $z_{i}$ is the energy fraction, and $\theta_{ij}$ denotes the angular distance. The choice of $z_i$ and $\theta_{ij}$ depends on the specific collider and the problem. For the subsequent discussion, the following choices are made \cite{Komiske:2017aww}:
\begin{align}
z_i &= \frac{p_{T,i}}{p_{T,J}},\hspace{2cm} p_{T,J} = \sum_{i=1}^M p_{T,i}\\
\theta_{ij} &= (\Delta y_{ij}^2 + \Delta \phi_{ij}^2)^{\beta/2}
\end{align}
where $\Delta y$ and $\Delta \phi$ denote the difference in rapidity and azimuth of the two particles, the exponent $\beta$ is fixed at 0.5. To determine the effectiveness of EFPs in discriminating top fat jets from quark/gluon-initiated jets, SM top pair and dijet events are generated using Pythia 8.226 \cite{Sjostrand:2014zea}. The final state visible particles are clustered into $R=0.8$ anti-$k_T$ jets. The leading fat jet in an event satisfying 500 GeV $<p_T<550$ GeV and $|\eta| < 1.7$ are kept for the final analysis. All EFPs up to degree $d\le 7$ are computed for these events. These EFPs are used as input in a Fisher's linear discriminant \cite{Fisher:1936et} as well as a three-layered DNN. Additional classifiers based on the N-subjettiness variable and jet images are also designed to compare the classification performance. We direct the interested readers to reference \cite{Kasieczka:2015jma} for a complete description of these classifiers. Here, we present the performance of the classifiers mentioned above in Figure \ref{fig:PFP}.

\section{Image-based Classifiers}
\label{sec:CNN}
\subsection{Image representation of jets and Preprocessing}
\label{sec:CNN-preprocess}
Jet image, like the images we see in our day-to-day lives, is a grid of numbers. Each grid cell is called a pixel, and the number associated with each pixel is called the pixel intensity. Pixels in jet images can have various sources. The simplest and most common source is the calorimeter tower, defined as the sum of energy deposits in a given rapidity-azimuthal angle bin of the Electromagnetic and hadronic calorimeter. Another source can be the topoclusters \cite{ATLAS:2016krp}, i.e., the clusters of calorimeter cells. The topoclusters help as most jet images are sparse with many empty pixels, and combining these empty pixels helps build a smaller image that requires a simpler neural network architecture. Finally, tracks can also be used as a source of pixels. Now, tracks have a much finer granularity than the towers and hence can provide much better recognition of physical features present in a jet. At the same time, analyzing a large image is computationally expensive. Therefore, a bunch of tracks in a given $\eta - \phi$ direction are usually grouped to form simpler jet images. The CMS collaboration also uses ParticleFlow \cite{CMS:2017yfk} algorithms to identify calorimeter energy deposits originating from charged particles (i.e., with associated charged tracks). These particle flow candidates can also be used to construct jet images. \\

Jet images can be uni-layered (grayscale images) or multi-layered (colored images). Colored images usually contain several layers originating from the sources mentioned above. Each layer corresponds to a different part of the detector, providing better insight into the jet content. Construction of such images can be complicated as different parts of the detector can have different resolutions, and all layers of the images must be of similar shape. In such situations, the pixels coming from the high-resolution source are usually grouped to match the resolution of the other source. This may lead to the loss of important information. \\

After the images are formed, they are passed through a series of pre-processing steps before being passed into any neural network. These preprocessing steps remove any redundant information in the images, thereby helping the neural network learn important characteristics of the jet more efficiently. Apart from stabilizing the training process, it improves the classifier's performance by simplifying the training data. However, these preprocessing steps can sometimes lead to the loss of important information and may hamper the classifier's performance.\\ 

Four common pre-processing steps are usually employed in jet image studies: Translation, Reflection, Rotation, and Normalization. In translation, all the jet constituents are shifted so that the constituent with the highest transverse momentum lies at the origin of the ($\eta,\phi$) coordinate system. 
Now, translation along the $\phi$ direction does not cause any additional trouble as there is no preferred direction in the transverse plane, and the physics should remain the same when moving the system as a whole along $\phi$. Translation in the $\eta$ direction, on the other hand, is equivalent to a longitudinal boost along the beam axis. Therefore, if quantities like energy deposit are used as pixel intensities, which are not invariant to longitudinal boost, important jet characteristics, like the invariant mass, may change. Therefore, quantities like transverse momentum or transverse energy are usually preferred. In rotation, the entire jet is rotated such that the second highest pt constituent or the principal component of the pixel intensity distribution becomes vertical. It is based on the reasoning that the radiation inside a jet is symmetric and, therefore, important characteristics of the jets should not change when the jet is rotated. However, this argument is not entirely true. The main difficulty arises due to pixelization. With square pixels at hand, any rotation by an angle other than $\pi/2$ will distort the jet. Researchers try to avoid this problem by using interpolation techniques to minimize this information loss. In reflection, the jet image is flipped so that the part of the image with a higher energy deposit lies to the left/right. Sometimes, a second flip is performed to make one of the corners the quadrant with the highest energy deposit. Finally, in normalization, the pixels are normalized either naively, $a_i\rightarrow a_i/\sum_i a_i$, or through the $L^2$ normalization,  $a_i\rightarrow a_i/\sum_i a_i^2$, here $a_i$ is the pixel intensity. Note that the $L^2$ normalization does not preserve the invariant mass. After preprocessing, the jet image can be used to train or test the performance of a classifier.\\
\subsection{Convolutional Neural Network}
Though image representation of jets can be used as inputs to several NN-based architectures (See \cite{Baldi:2016fql, Oliveira2017LearningPP} for the use of locally connected layers on image data), the most common and efficient ones are CNNs. This popularity is due to the translational invariance of convolution operations and the weight-sharing technique employed by CNNs(see the discussion below).\\

CNNs use convolutional filters/grids containing weights that scan the input image. This scanning involves performing the inner product between different patches of the image (of the same dimension as the filter) and the filter. The output of this inner product serves as a new pixel for the output image. Once an inner product is performed, the filter is moved by certain units called stride, producing another pixel. This process is repeated till the full image is covered. The output pixels are combined to produce the response map. For uni-layered images, these filters are two-dimensional. On the other hand, for colored images, three-dimensional filters with depth equal to the number of image layers are used. The output of each convolution operation is one single layer of the response map; the response map's depth is determined by the number of filters used. The use of the same filter for different parts of the image is the weight-sharing mechanism mentioned earlier, and it is highly efficient in reducing the number of trainable parameters. After the convolution operation, the response map is passed through a non-linear activation to produce the activation map.\\

The convolution operation produces output images with sizes smaller than the input. This can cause trouble in constructing very deep networks. To overcome this, padding is performed where some empty pixel layers are added to the edges of the input image. The size reduction is usually achieved with the help of some pooling operation, usually max pooling or mean pooling. The pooling operation resembles the convolution, where a squared grid is moved across the image to produce the final output. However, the difference is the pooling filter does not contain any weights. Its only job is to combine the pixels through the pooling operation. After many such convolution and pooling operations, the output image is linearised and is passed through a series of fully connected layers called the decoding network to produce the final output (either a classification score or a regression value). \\

Another problem commonly encountered in deep CNNs is the so-called vanishing gradient problem (VGP). The updation of network parameters (weights and biases) in NNs happens through backpropagation, which implements the chain rule to determine the gradients of the loss function with respect to the network parameters. The network uses these gradients with the learning rate (which determines the step size) to update the parameters that minimize the loss function. In very deep NNs, the updating of parameters in initial nodes involves a product of a large number of derivatives. This may generate vanishing gradients, where gradients in earlier layers become very small during training. This phenomenon can slow down the learning process or hinder the training of deep networks. While parameter updating typically continues in neural networks despite vanishing gradients, they can affect convergence speed and overall training effectiveness. Over the years, several techniques have been developed to address the vanishing gradient problem (VGP) in neural networks. A brief discussion on some of these techniques is in order.
\begin{itemize}
	\item Unitary matrices for weight initialization \cite{le2015simple, jing2017tunable, saxe2014exact}: Since the eigenvalues of a unitary matrix are unimodular and thus can be raised to arbitrary power without vanishing, using unitary matrices helps stabilize the training process.
	\item Artificial Derivatives \cite{9336631}: For sigmoid and ReLU activation functions, artificial derivatives can amplify the loss function's derivatives when they approach zero, thereby addressing the VGP effectively.
	\item Specialized RNN architectures: Long short-term memory (LSTM) networks \cite{10.1162/neco.1997.9.8.1735} and gated recurrent units (GRUs) \cite{cho2014properties} uses a gating mechanism to maintain and control the flow of gradient over long sequences, facilitating stable learning.
	\item Weight initialization methods: Using methods like Xavier \cite{GloBen10Understanding} or He \cite{he2015delving} appropriate to the activation function ensures proper gradient scaling during backpropagation.
	\item Gradient clipping: This technique limits the gradients' magnitude, preventing them from becoming too large or too small.
	\item Batch Normalization \cite{ioffe2015batch}: By normalizing the inputs of each layer, batch normalization improves the stability and convergence during training.
	\item Residual connections \cite{he2015deep}: The input to the network is passed to the output through a skipped connection. This means instead of the usual output $f(x)$ (x being the input and f being a combination of operations like pooling, convolution, etc), the new output becomes $x + f(x)$. Note that sometimes the input has to undergo additional operations to change its size to be combined with $f(x)$.
\end{itemize}

Out of these techniques, appropriate weight initialization, batch normalization, gradient clipping, and residual connections are commonly employed with CNNs.

\subsection{CNN architectures}
This section discusses some of the state-of-the-art CNN architecture for top tagging. We plan to present our study in chronological order, mainly focusing on five imaged-based analyses: DeepTop \cite{Kasieczka:2017nvn}, Upgraded Deeptop \cite{Macaluso:2018tck}, ResNext-50 \cite{Kasieczka:2019dbj}, CapsNet \cite{Diefenbacher:2019ezd}, and Bayesian networks \cite{Bollweg:2019skg}. In all the studies discussed below, the SM top pair production is used to generate the signal images, and QCD di-jet production provides the background sample unless mentioned otherwise.\\

\begin{figure}[!htb]
	\centering
	\includegraphics[width=0.8\columnwidth,width=0.35\textheight]{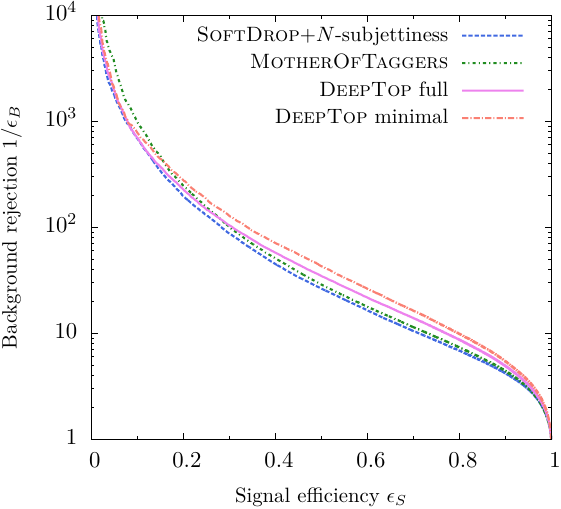}
	\caption{The ROC curves for the minimal and full version of the DeepTop tagger compared with the HLF-based Softdrop + N-subjettiness and MotherOfTaggers \cite{Kasieczka:2017nvn}.}
	\label{fig:deeptop}
\end{figure}

The DeepTop CNN was originally introduced in reference \cite{Kasieczka:2017nvn}. It uses single-layered top images based on the calorimeter energy deposit as input. Signal and background fat jets in the transverse momentum range of 350-450 GeV and $|\eta| < 1$ are considered for the analysis. After detector simulation, the anti-k$_T$ fat jets with R=1.5 are reconstructed, and the calorimeter towers are used for image construction with the transverse energy serving as pixel intensity. The images are passed through a series of pre-processing steps: pixelization, translation, rotation, reflection, and scaling. We discussed translation, reflection, and rotation in the previous sections. In pixelization, the image is divided into $40\times 40$ pixels, while scaling ensures all pixel intensities are in the range of 0 to 1. Two types of images are considered for the final analysis: one that does not incorporate the reflection and rotation preprocessing steps (DeepTop minimal) and another that does (DeepTop full). To better understand the effectiveness of the DeepTop tagger, its performance is compared with two baseline BDT-based taggers called SoftDrop + N subjettiness and MotherOfTaggers \footnote{Mainly consisting of the HEPTopTagger\cite{Plehn:2010st, Kasieczka:2015jma} variables.}. The architecture and model implementation details can be found in reference \cite{Kasieczka:2017nvn}; we only showcase their final results in figure \ref{fig:deeptop}. To summarise their results, the CNN-based taggers perform slightly better than the High-level feature-based taggers, proving their effectiveness. Between the two CNN taggers, the DeepTop minimal performs slightly better than the DeepTop full. This behavior can be ascribed to the loss of information, as discussed in section \ref{sec:CNN-preprocess}, in the additional pre-processing steps used in the DeepTop full images.\\

\begin{figure}[!htb]
	\centering
	\includegraphics[width=0.45\columnwidth]{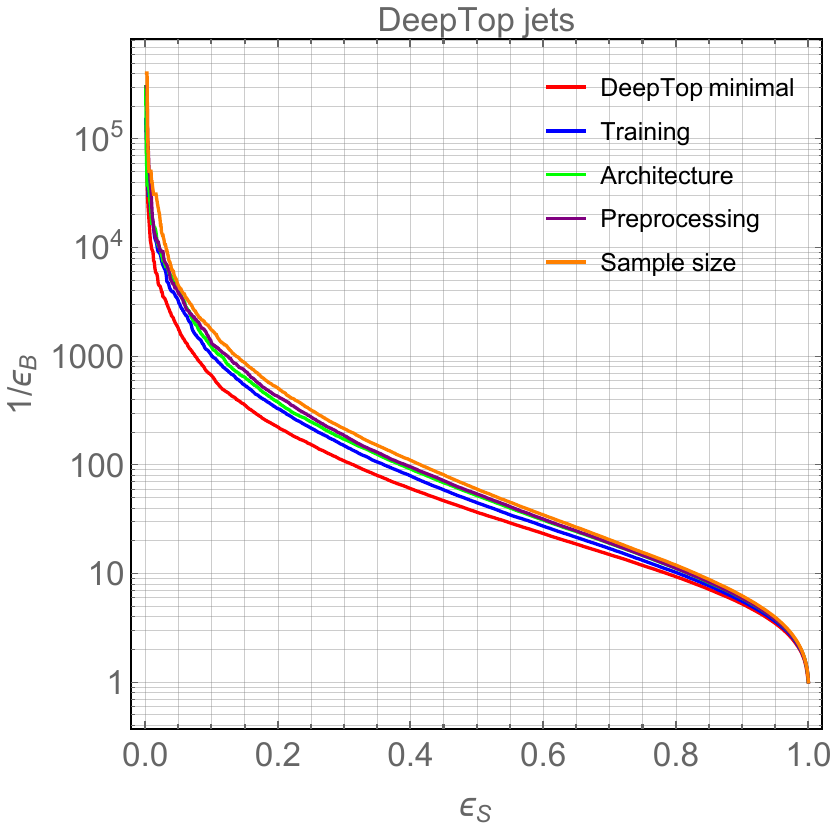}
	\includegraphics[width=0.45\columnwidth]{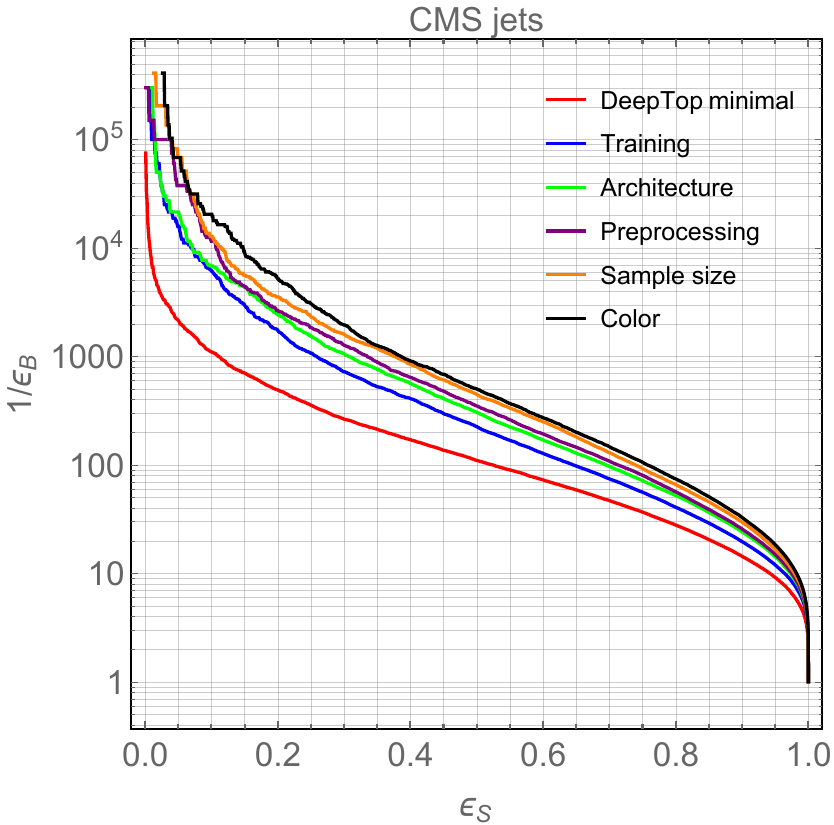}
	\caption{The ROC curves for the DeepTop jets(left) and CMS jets (right) samples demonstrating the gradual improvement in performance as modifications are incorporated to the minimal DeepTop model \cite{Macaluso:2018tck}.}
	\label{fig:updateddeeptop}
\end{figure}

An upgraded version of the DeepTop analysis was presented in reference \cite{Macaluso:2018tck}. The authors made several improvements to the previous analysis \cite{Kasieczka:2017nvn}. These involve a change in loss function from Mean Square Error to Cross Entropy (better suited for classification problems), a change of optimizer from Stochastic Gradient Descent to AdaDelta \cite{zeiler2012adadelta} (to take advantage of the insensitivity of the latter to noisy gradients), introducing a learning rate scheduler, improved architecture with an increased number of nodes and more feature maps, change in the pre-processing \footnote{Instead of creating the image pixels before the preprocessing step they moved it to the end. As a result, the preprocessing steps can now utilize the fine granularity of the tracks. They also performed a two-fold flipping, first left-right, then up-down.}, increased sample size, and introduced multi-layered / color images. This analysis also considered two types of training data. The first set resembles the sample of the DeepTop analysis with top/QCD jets in the $p_T$ range of 350-450 GeV, reconstructed with R=1.5 anti-k$_T$ jets after detector simulation. These jets are matched with a truth level top by demanding $\Delta R(t,j) < 1.2$. This sample only uses single-layered images generated from the calorimeter and utilizes the transverse momentum as pixel intensity. After preprocessing, the images are pixelized into a $40\times 40$ grid. The second dataset (CMS jets) uses top/QCD jets in the $p_T$ range 800-900 GeV, reconstructed with a $R=0.8$. Apart from the matching condition of $\Delta R(t,j) < 0.6$, an additional merging condition is demanded that ensures the presence of all the top decay products (at parton level) inside the top jet. The second dataset also employs color images with four layers originating from calorimeter $p_T$ of the neutral constituents, per-pixel track $p_T$, per-pixel track, and muon multiplicity. The model implementation and network architecture we direct the interested reader to reference \cite{Macaluso:2018tck}. We only present their final results for the two datasets in figure \ref{fig:updateddeeptop}. The figure also demonstrates the gradual improvement in performance from the previous DeepTop tagger as the new modifications are sequentially incorporated.\\

\begin{figure}[!htb]
	\centering
	\includegraphics[width=0.8\columnwidth]{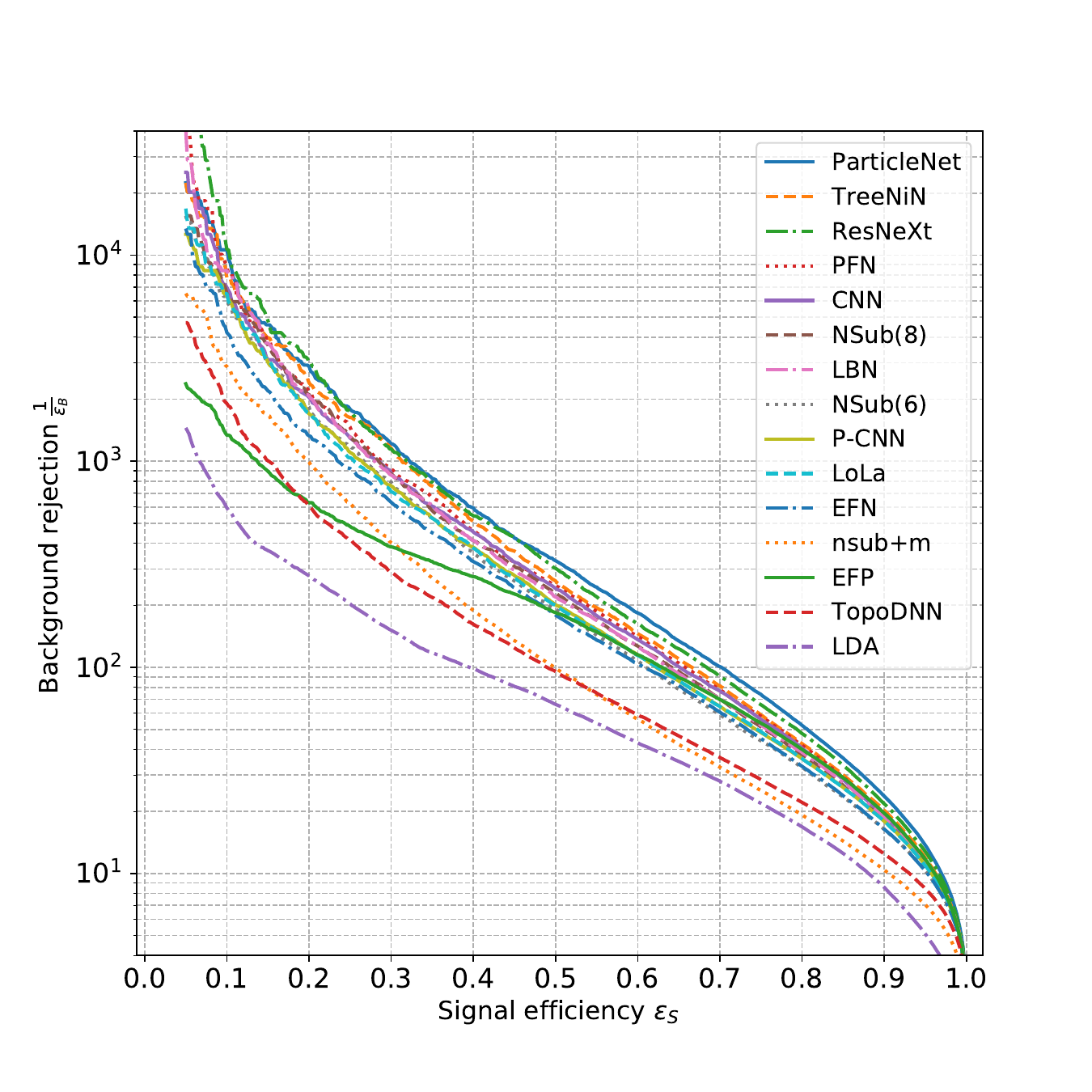}
	\caption{ROC curves showcasing the performance of ResNext and several other image-based and four-vector-based classifiers \cite{Kasieczka:2019dbj}.}
	\label{fig:resnet}
\end{figure}

The ResNeXt architecture was implemented in reference \cite{Kasieczka:2019dbj} for top tagging. The model used was a smaller version of the original ResNeXt-50 \cite{xie2017aggregated} architecture, with the number of channels in all layers except the first reduced by a factor of four. This modification is due to the smaller size of the jet images used. The jet images are reconstructed from particle flow candidates collected in a $64 \times 64$ grid. Reference \cite{Kasieczka:2019dbj} studied top/QCD jets in the $p_T$ range of 550-650 GeV, reconstructed as R=0.8 anti-k$_T$ jets after dedicated detector simulation. Both matching and merging requirements are imposed to ensure the selection of properly reconstructed fat jets. The performance of ResNeXt was compared with several other DNN and GNN-based architectures. We present their final result in figure \ref{fig:resnet}.\\

\begin{figure}[!htb]
	\centering
	\includegraphics[width=0.45\columnwidth]{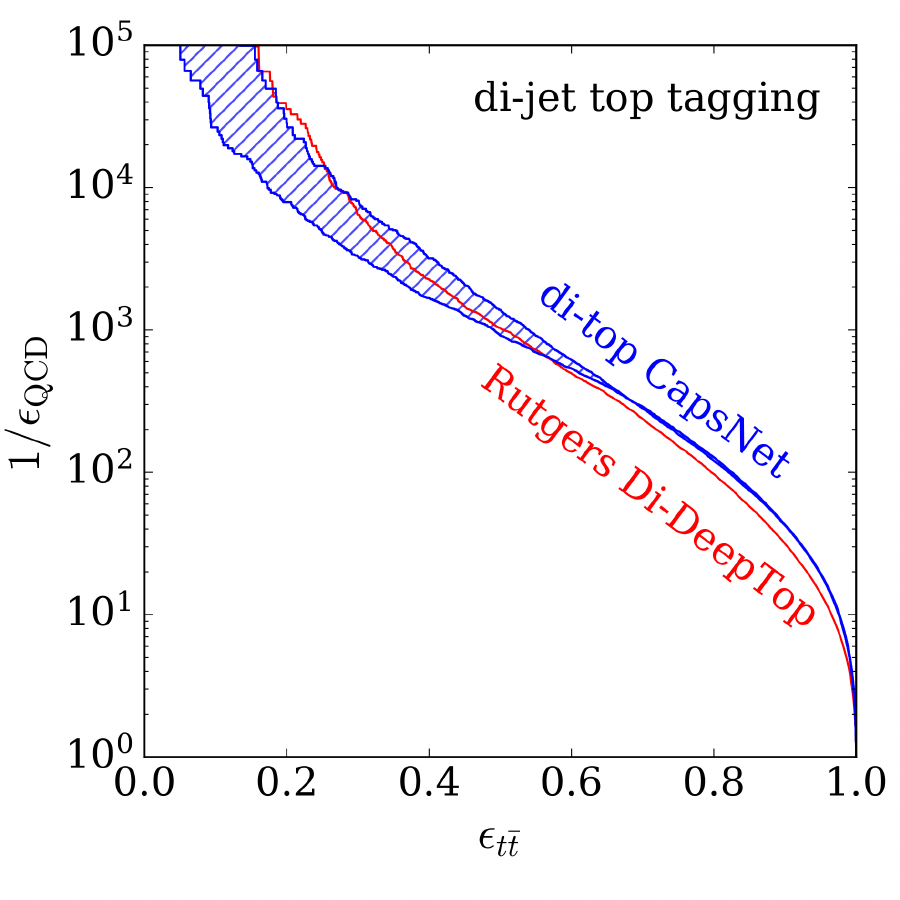}
	\includegraphics[width=0.45\columnwidth]{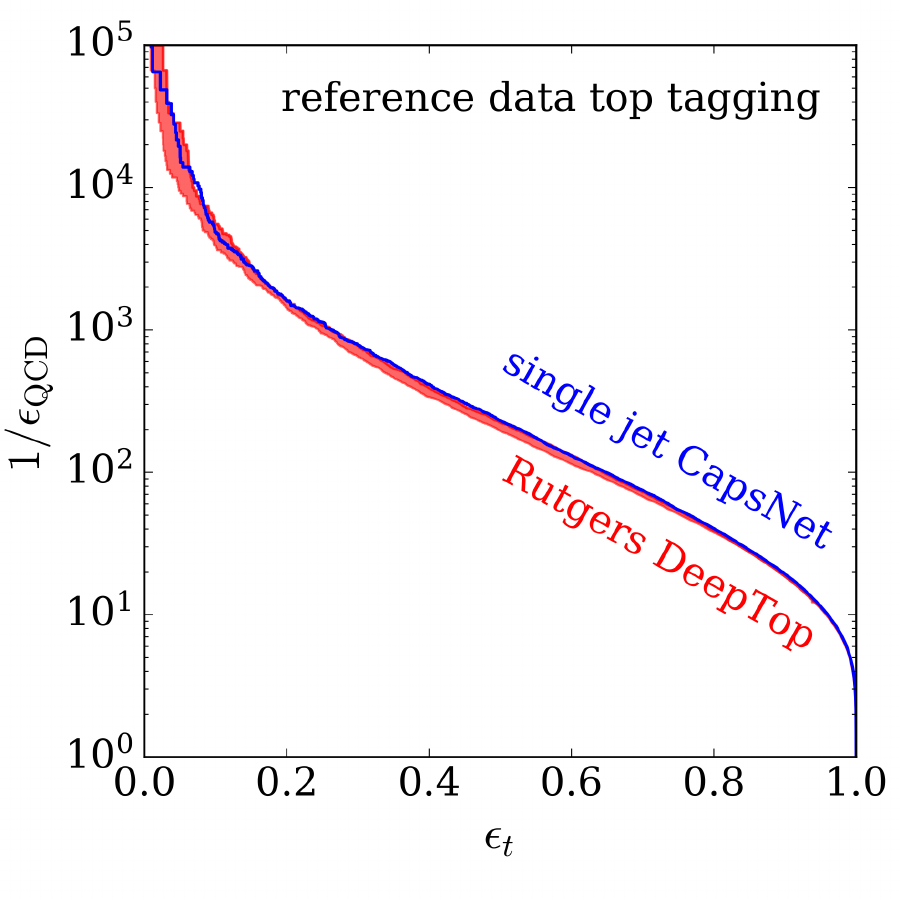}
	\caption{The ROC curves for CapsNet and Rutgers DeepTop CNN for di-top tagging (left) and single top tagging (right) \cite{Diefenbacher:2019ezd}. The blue-shaded region represents the uncertainty stemming from the use of different estimators to build the ROC curve.}
	\label{fig:capsnet}
\end{figure}

The application of Capsule Networks for collider analysis was first demonstrated in reference \cite{Diefenbacher:2019ezd}. The model implementation follows the original paper \cite{sabour2017dynamic}. Though the analysis \cite{Diefenbacher:2019ezd} mainly focuses on the application of capsules for event-level analysis in separating a resonance decaying into a top pair from continuum top and dijet backgrounds and emphasizing the usefulness of capsules for overlying images, they have also studied the use of CapsNet for di-top and single top tagging. The usefulness of capsules for event-level analyses stems from their ability to learn the geometric position and orientation of objects in an image. In CNNs, the final classification/regression is done by linearising the images and using a decoding layer to generate the network prediction. CapsNet converts the images into capsule vectors in signal/background feature space. For instance, a 12-layer $8\times 8$ image in CNN gets transformed into a linear array of $12\times 8\times 8$ numbers, while in a CapsNet, it can transform into $8\times 8 = 64$ capsules of dimension 12 each. These capsules are further processed till they reach the final layer, with the number of capsules equal to the number of classes under study. The length of these capsules encodes how likely it is for the event in that particular class. More details on the model implementation and signal/background samples can be found in ref \cite{Diefenbacher:2019ezd}. In figure \ref{fig:capsnet}, we present the results of ref \cite{Diefenbacher:2019ezd} for di-top and single-top tagging. The di-top samples are generated following the process $p p \rightarrow Z^{\prime} \rightarrow t \Bar{t}$, and the corresponding background comes from QCD dijet events. After detector simulation, the top jets are reconstructed as $R=1.0$ C/A jets. Jets with $p_T > 350$ and $|\eta| < 2.0$ are considered for further analysis. The whole calorimeter energy deposit is converted into size $180\times 180$ images with transverse energy as the pixel intensity. For the single top analysis, the public dataset \cite{KasPleThRu19} was used. For the sake of comparison, the analysis also considered the performance of the DeepTop tagger \cite{Macaluso:2018tck} for both tasks. As can be seen from Fig. \ref{fig:capsnet}, both DeepTop and CapsNet demonstrate comparable performance. The blue-shaded region in Fig. \ref{fig:capsnet} represents the uncertainty stemming from the use of different estimators. In conventional CNNs with a softmax activation function in the final layer, the output of the signal and background neurons are not independent, and one usually uses the output of the signal neuron as an estimator to build the ROC curve. However, the same is not true for the CapsNet. Capsules are vectors that are designed to encode the geometric position and orientation of objects in an image. Consequently, the signal and background capsules carry independent information. This gives one the freedom to try out different estimators and test their effect on the tagging performance. Ref.~\cite{Diefenbacher:2019ezd} used two such estimators (see Ref.~\cite{Diefenbacher:2019ezd} for detailed discussion), and the region in blue represents the area between the ROC curves resulting from these two estimators. It is important to note that the choice of the estimator affects the tagging performance of realistic training.\\

\begin{figure}[!htb]
	\centering
	\includegraphics[width=0.8\columnwidth]{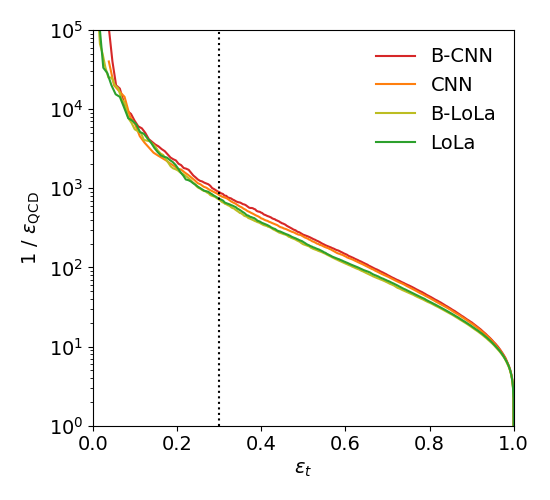}
	\caption{ROC curves for the Bayesian and deterministic versions of DeepTop and DeepTopLoLa taggers \cite{Bollweg:2019skg}.}
	\label{fig:bayesian}
\end{figure}

The Bayesian version of the DeepTop tagger was originally introduced in reference \cite{Bollweg:2019skg}. Bayesian neural networks (BNNs) have the advantage that in addition to the network score, they also provide the score distribution, which can be used to estimate the error band on the score. BNNs help determine statistical uncertainties due to the limited size of the data sample and other systematics originating from pile-up interactions and jet energy scale. The goal is to determine the posterior distribution of the network weights given an assumed prior. The Kullback-Leibler divergence is used to estimate the approximate shape of the posterior from the training data. Once the posterior distribution is estimated, it can be used to determine the mean network output and the associated error. In reference \cite{Bollweg:2019skg}, top/QCD jets in the $p_T$ range 550-650 GeV are used to check the performance of Bayesian DeepTop and compare it with the corresponding deterministic version. After reconstruction, the transverse energy of the calorimeter cells is used to construct the jet images. These images are passed through the updated preprocessing steps suggested in reference \cite{Macaluso:2018tck}. In addition to the image-based DeepTop tagger, the Bayesian version of the four-vector-based DeepTopLoLa tagger \cite{Butter:2017cot} was also considered. We present the ROC curves for the Bayesian and deterministic versions of DeepTop and DeepTopLoLa taggers in Figure \ref{fig:bayesian}.

\section{GNN}
\label{sec:GNN}
\subsection{Graph representation of jets}
In the previous section, we discussed the remarkable success of CNNs in separating boosted jets originating from the hadronic decay of tops from those originating from light quarks and gluons. The reason for this success is twofold: firstly, CNN architecture respects translational symmetry and can identify patterns in different parts of the image. Second, CNN's weight-sharing mechanism helps drastically reduce the number of trainable parameters. This suggests that NN architectures that respect the intrinsic symmetry of the problem can not only excel in performance but can also provide a more economical network design. \\

Returning to the case at hand, the constituents of a fat jet originate from the showering, fragmentation, and hadronization of the initial parton. We observe these constituents as charged tracks and calorimeter energy deposits at detectors. Though most analyses try to order these constituents by their transverse momentum, energy, or angular position, no theoretical reasoning supports this assumption. Therefore, architectures that respect this inherent unordered nature of jet constituents possess the potential to capture intricate jet-level features adeptly. This calls for the graph representation of fat jets, facilitating the development of permutation-equivariant architectures and enabling the modeling of intricate interrelations among jet constituents. Graph Neural Networks (GNNs) are architectures designed to learn functions on these graphs. For a comprehensive understanding of GNNs, we encourage the interested reader to consult reference \cite{Shlomi:2020gdn,Thais:2022iok}. For our discussion, we only noted some key features of graphs.\\

A graph is a collection of nodes and the pairwise relationship between these nodes (edges). In the case of a fat jet, these nodes can represent the jet constituents, with the constituent four vectors playing the role of the node coordinates in Minkowski space. Each node can also be characterized by node-specific features like the constituent mass, charge, or ID. In addition, we can also associate each node with some global features characteristic of the fat jet. In other words, we can represent the graph node as $f_i = x_i \oplus h_i \oplus g_i$, with the $x_i$ representing the constituent four-vector, $h_i$ the constituent features and $g_i$ the global features. Similarly, we can represent the graph as G(F, E) with F as the vertices and E as the edges. The action of a GNN on the graph can be described as a sequence of message-passing operations where, at each step, the information from the node neighborhood is collected, and the node characteristics are updated. Quantitatively, we can represent the lth message passing operation as 
\begin{align}
m^{l+1} &= \sum_{j\in[N(i)]} \phi_1(f_i^l,f_j^l,e_{ij}^l)\\
f_i^{l+1} &= \phi_2(f_i^l,m^{l+1})
\end{align}
Here, the sum runs over the neighborhood of node $i$, $e_{ij}$ denote the edge function. The function $\phi_1$ and $\phi_2$ are usually replaced with neural networks. Note that the first message-passing layer transforms the node features from physical to latent space. The subsequent definition of the node neighborhood relies on the latent space representation of the graphs.

\subsection{GNN architectures}
This section will discuss some of the leading GNN architectures for top tagging. Despite the presence of many well-to-do and physically motivated architecture, we have decided to focus our discussion on five models: PFN \cite{Komiske:2018cqr}, ParticleNet \cite{Qu:2019gqs}, LGN \cite{Bogatskiy:2020tje}, LorentzNet \cite{Gong:2022lye}, and PELICAN \cite{Bogatskiy:2023nnw, Bogatskiy:2022czk, Bogatskiy:2023fug}. Each of these models has used the public dataset \cite{KasPleThRu19} to access the classifiers' performance, giving us a common ground to compare their performance.\\

\begin{figure}[!htb]
	\centering
	\includegraphics[width=0.8\columnwidth]{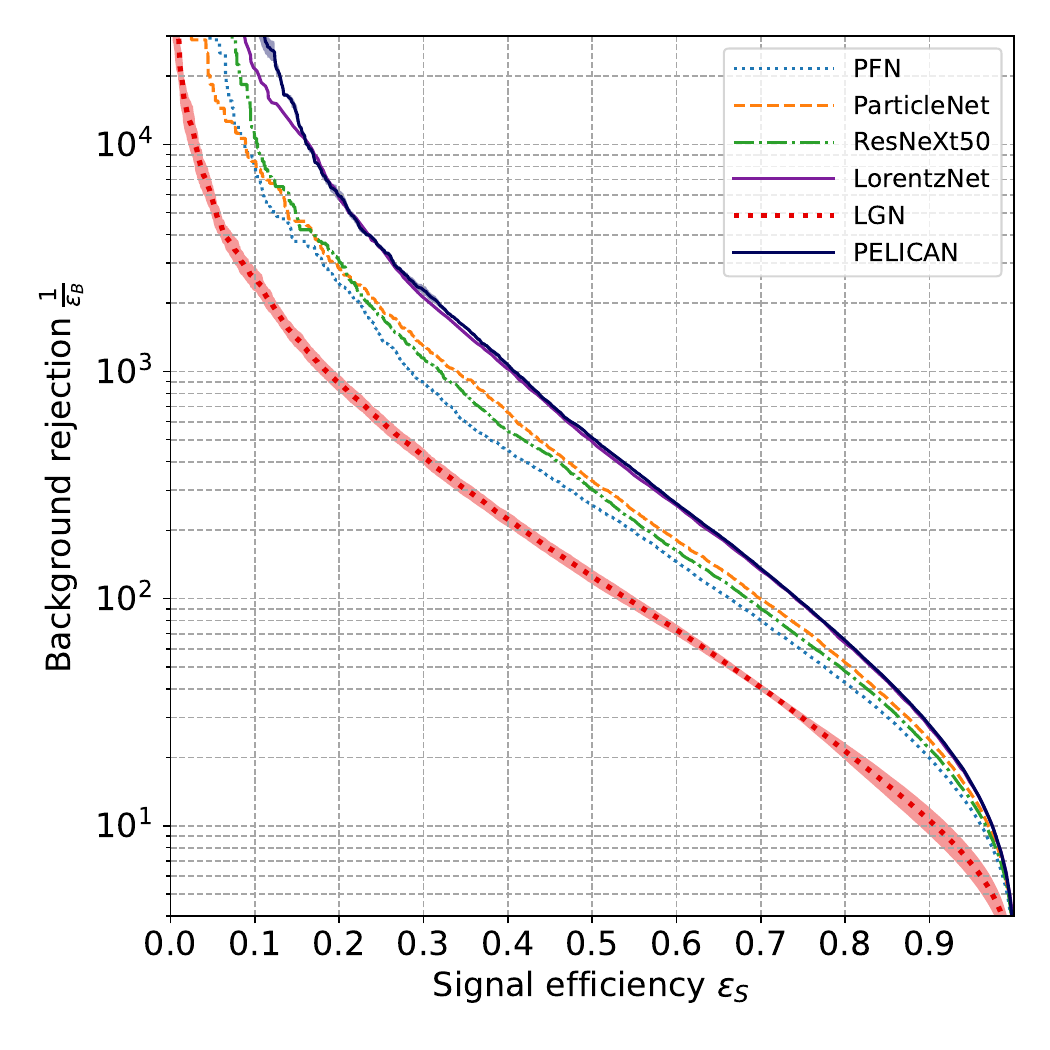}
	\caption{ROC curves for the GNN classifiers \cite{Bogatskiy:2022czk}.}
	\label{fig:gnn}
\end{figure}

The Particle Flow Network (PFN) \cite{Komiske:2018cqr} operates on a point cloud representation of jet constituents, disregarding any inherent ordering among them. The theoretical basis of PFN is the Deep Sets theorem, which suggests that any observable associated with a jet can be approximated by \cite{Komiske:2018cqr}:
\begin{equation}
O({p_1,....p_M}) = F(\sum_{i=1}^M \phi(p_i))
\end{equation}
Here, $p_i$ ($i$ = 1...M (the number of constituents) ) represents some characteristics of the jet constituents like their four-momentum, mass, identification, etc. $\phi$ act on each particle transforming its feature to the latent space. The summation over the particle label ensures permutation equivariance and converts the per-particle latent representation into the latent representation of the event. Finally, the map $F$ converts the event latent representation into the final observable. In PFN, these functions $\phi$ and $F$ are realized through neural networks. The model performance is evaluated on the public top-tagging dataset \cite{KasPleThRu19}. Before passing through the network, the dataset undergoes some preprocessing where the jet is first centered in the $\eta-\phi$ plane, the $p_T$ of the constituents are normalized, and undergo rotation and reflection. For more details on the model implementation and dataset used, we refer the interested reader to the original paper \cite{Komiske:2018cqr}. Here, we present their results in Figure \ref{fig:gnn} and Table \ref{tab:GNN}. Note that the architecture discussed here is named PFN-r.r. in reference \cite{Komiske:2018cqr}. In their work, PFN denotes a model excluding reflection and rotation preprocessing steps. However, as evidenced in reference \cite{Komiske:2018cqr}, these preprocessing steps help in enhancing the classifier's performance. Finally, we also want to mention that reference \cite{Komiske:2018cqr} also implements an infra-red and collinear (IRC) safe version of the network called Energy Flow Network (EFN). Though the network makes more physical sense \footnote{IRC safety ensures that the network output (the so-called {\it topness} of the jet) is invariant under soft and collinear emissions. Such observables are calculable within the paradigm of perturbative QCD (see the discussion in Ref.~\cite{Konar:2021zdg}) and thus are more sensible.}, it performs slightly less on the top-tagging dataset. We refrain from discussing EFN here and urge the interested reader to consult \cite{Komiske:2018cqr}.\\

Like PFN \cite{Komiske:2018cqr}, ParticleNet \cite{Qu:2019gqs} also employs the point cloud representation of jet constituents (particle cloud). However, unlike PFN, which follows the Deep Sets approach, ParticleNet implements edge convolution \cite{edgeconv3326362} that helps the architecture utilize the local structure of the constituents. The edge convolution operation is inspired by the weight-sharing and hierarchical learning features of CNNs. However, unlike images, the point clouds can have uneven shapes and do not have a grid-like representation. This makes it difficult to construct local patches for the convolution kernel to operate. Edge convolution solves this problem by using the k nearest neighbors to construct local patches in a point cloud. Analytically, for a given point $x_i$, the operation of edge convolution results in\cite{Qu:2019gqs}:
\begin{equation}
x_i^{\prime} = \overset{k}{\underset{j=1}{\Box}} h_{\Theta}(x_i,x_{i_j} - x_i)
\end{equation}
Here, the sum runs over the k-nearest neighbors of the point $x_i$, i.e. $\{x_{i_1}.. x_{i_k}\}$. $\Theta$s are some learnable parameters parametrizing the function $h_{\Theta}$, and $\square$ denotes the aggregation operator, designed to respect permutation equivariance (ParticleNet uses a mean aggregator). ParticleNet implements $h_{\Theta}$ using neural networks with parameters shared across the edges. The main blocks of ParticleNet are the EdgeConv blocks. These blocks first determine the k nearest neighbors of a particle using their position in the rapidity-azimuth plane. The edge features are then passed through the EdgeConv operation. ParticleNet also uses residual connections to pass the input features to the output of the EdgeConv operation. The combined output is passed through the subsequent EdgeConv block. Note that the second EdgeConv block utilizes the latent representation of the point cloud while determining the k nighest neighbors. In other words, the point clouds are dynamically updated, making ParticleNet a Dynamic Graph Convolutional Neural Network (DGCNN). ParticleNet implements three such EdgeConv blocks, each with k=16. After the EdgeConv blocks, the output undergoes a global average pooling followed by a decoding layer to produce the predictions. For complete details on the architecture, we urge the interested reader to consult the original paper \cite{Qu:2019gqs}. The model performance is evaluated on the public top tagging dataset \cite{KasPleThRu19}, and we present their results in Figure \ref{fig:gnn} and Table \ref{tab:GNN}.\\

\begin{table}[htb!]
	\centering
	\begin{tabular}{l@{\hspace{2mm}}|l@{\hspace{2mm}}|l@{\hspace{2mm}}|l@{\hspace{2mm}}|r}
		\toprule
		Architecture    &   Accuracy    &   AUC         &   $1/\epsilon_B$  &   \# Params \\
		\midrule
		LGN             &   0.929(1)    &   0.964(14)   & 424 $\pm$ 82      &   4.5k    \\
		PFN             &   0.932       &   0.982       &   891 $\pm$ 18    &   82k     \\
		ResNeXt         &   0.936       &   0.984       &   1122 $\pm$ 47   &   1.46M   \\
		ParticleNet     &   0.938       &   0.985       &   1298 $\pm$ 46   &   498k    \\
		LorentzNet      &   0.942       &   0.9868      & 2195 $\pm$ 173    &   220k    \\     
		PELICAN         &   0.9425(1)   &   0.9869(1)   & 2289 $\pm$ 204    &   45k     \\   
		\bottomrule
	\end{tabular}
	\caption{Performance and architectural complexity of different GNN top taggers \cite{Bogatskiy:2022czk}. The results are calculated from the average of several training runs (the number of runs varies across networks, see Ref.~\cite{Bogatskiy:2022czk}) with random network initialization. The numbers in the parenthesis represent the uncertainties over these runs.}
	\label{tab:GNN}
\end{table}

Lorentz invariance is the fundamental symmetry of space-time governing elementary particle interactions. NN architecture respecting this symmetry can provide a relatively simple and physically interpretable design. Lorentz Group Network (LGN) \cite{Bogatskiy:2020tje} is based on the theory of finite dimensional representation of the Lorentz group and demonstrated for the first time the usefulness of including Lorentz equivariance in the construction of efficient GNN architectures with significantly fewer trainable parameters. LGN is based on the G-equivariant universal approximation theorem \cite{yarotsky2018universal} that suggests that any equivariant map between two completely reducible representations of a lie group G can be realized through NNs using vector activations belonging to finite-dimensional representations of the group G. The permutation equivariant and Lorentz equivariant architecture is built by stacking several Clebsch-Gordon layers \cite{Bogatskiy:2020tje} on top of one another that perform CG decompositions on the activations of the previous layers. It first performs tensor products representing self-interactions and interactions among different particles. The CG operator acts upon these tensor products to decompose them into irreducible representations of the Lorentz group. Finally, an equivariant learnable operator mixes these decompositions to form the channels of the next layer. Apart from the CG layer, the architecture also includes an input layer, several MLP layers, and an Output layer. For the details of the model implementation, see \cite{Bogatskiy:2020tje}. The performance of the model is evaluated on the public dataset \cite{KasPleThRu19}, and we present the model performance in Figure \ref{fig:gnn} and table \ref{tab:GNN}.\\

LorentzNet \cite{Gong:2022lye} is another permutation equivariant and Lorentz equivariant GNN architecture. Like the CG layers in LGN \cite{Bogatskiy:2020tje}, the main building blocks of LorentzNet are called Lorentz Group Equivariant Blocks (LGEB). Their role is to define the edges of the graph and update the node coordinates and the node embeddings while respecting Lorentz equivariance. If we denote the node embeddings of the lth layer as $h_i^l$ (i=1,2...N) and the corresponding node coordinates in the Minkowski space as $x_i^l$ (I = 1,2...N), where N denotes the number of nodes then the action of the LGEBs can be summarised in three simple steps\cite{Gong:2022lye}:
\begin{itemize}
	\item Neighbourhood aggregation:
	\begin{equation}
	m_{ij}^l = \phi_e\left( h_i^l,h_j^l,\psi(||x_i^l-x_j^l||^2,\psi(\langle x_i^l,x_j^l\rangle))\right)
	\end{equation}
	where $\phi_e$ is a NN, $\psi(a) = sign(a)log(|a|+1)$ is introduced to normalize large entries, $||a||^2$ is the Minkowski norm, and $\langle a \rangle$ is the Minkowski inner product. LorentzNet does not assume any prior knowledge of interaction among the nodes; in other words, the graphs in LrentzNet are fully connected.
	\item Updating the node coordinates:
	\begin{equation}
	x_i^{l+1} = x_i^l + c \sum_{j \in [N]} \phi_x(m_{ij}^l).x_j
	\end{equation}
	Where the constant c is introduced to control the scale of $x_i^{l+1}$, $\phi_x$ is another NN, and the sum is over the neighborhood of the node $i$.
	\item Updating the node embeddings:
	\begin{equation}
	h_i^{l+1} = h_i^l + \phi_h\left( h_i^l, \sum_{j \in [N]} w_{ij} m_{ij}^l\right)
	\end{equation}
	Here, $w_{ij} = \phi_m(m_{ij})$ is introduced to code the significance of the edge between node $i$ and $j$, $\phi_m$ and $\phi_h$ are Neural Networks. 
\end{itemize}
Apart from the LGEBs, LorentzNet also contains an input/encoding layer that transforms the input node embedding scalars to the latent space and a final decoding layer that uses the output of LGEBs to generate Network predictions. The model's performance is tested on the publicly available dataset \cite{KasPleThRu19}, and we present the model's performance in Figure \ref{fig:gnn} and Table \ref{tab:GNN}.\\

So far, PELICAN \cite{Bogatskiy:2022czk, Bogatskiy:2023nnw} is the best-performing NN architecture for top-tagging. It is also a permutation equivariant and Lorentz equivariant GNN. Notably, PELICAN adeptly constructs a comprehensive set of Lorentz invariant functions, as per ref \cite{weyl1946classical}, from the four-momentum of jet constituents.  These invariants can be constructed from the pairwise dot products of the four vectors, and for PELICAN, these dot products serve as the primary inputs. To ensure permutation equivariance, PELICAN followed reference \cite{pmlr-v151-pan22a, corso2020principal} to construct the complete list of equivariant aggregators (15 to be exact) that can transform the input dot products. Following aggregation, PELICAN further refines the aggregated values through scaling, employing a factor of $(N/\Tilde{N})^{\alpha}$, where $\alpha$ represents a learnable parameter, N signifies the count of particles in the event, and $\Tilde{N}$ denotes a constant indicative of the typical number of particles expected in such events. The main building block of PELICAN is the equivariant block that contains the layers for message formation, followed by the aggregation block that applies the 15 aggregation functions discussed earlier. After five such equivariant blocks, the output passes through a decoding block that generates the model predictions. For a comprehensive understanding of the model architecture, interested readers are directed to reference \cite{Bogatskiy:2022czk}. Reference \cite{Bogatskiy:2022czk} also uses the public dataset \cite{KasPleThRu19} for the performance assessment of PELICAN. We present their results in Figure \ref{fig:gnn} and Table \ref{tab:GNN}.

\section{Top Quarks and Physics Beyond the Standard Model}
\label{sec:TopNP}
Attempts to alleviate various shortcomings of the SM---theoretical problems like the hierarchy problem, the flavour problem, the strong $CP$ problem, the vacuum instability problem, as well as experimental inconsistencies like the nonzero neutrino masses and mixing, the baryon asymmetry of the universe, the presence of cold dark matter (DM) in the universe, and various flavour anomalies---have led to a plethora of theories or models going beyond the SM. Examples of such theories or models are Grand Unification Theories (GUTs) \cite{Georgi:1974sy,Pati:1974yy,Fritzsch:1974nn,Gursey:1975ki,Achiman:1978vg,Ma:1986we,Hewett:1988xc}, supersymmetry \cite{Volkov:1973ix,Wess:1974tw,Wess:1974jb,Ferrara:1974pu,Salam:1974ig,Babu:1987kp}, wrapped extra-dimensions \cite{Randall:1999ee,Cheng:2001vd,Appelquist:2000nn,Burdman:2006gy}, technicolour models \cite{Dimopoulos:1981xc,Chivukula:1995gu,Malkawi:1996fs,Hill:2002ap}, little Higgs theories \cite{Weinberg:1972fn,Georgi:1975tz,Georgi:1974yw,Arkani-Hamed:2001nha,Schmaltz:2005ky} and theories featuring dynamical or elaborated spontaneous symmetry breaking \cite{Mohapatra:1974gc,Weinberg:1975gm,Senjanovic:1975rk,Beg:1977ti,Mohapatra:1977mj,Susskind:1978ms,Senjanovic:1978ev,Dimopoulos:1979es,Eichten:1979ah,Babu:1987kp,Ma:2010us,Hsieh:2010zr}. As briefly discussed below, such models introduces various new particles: extra gauge bosons, extra scalars, leptoquarks, vector-like quarks, {\it etc}. In many scenarios, these new particles can have preferential couplings to the third-generation fermions, in particular, the top quark. As such, at the LHC, their production and decay could result in various top-enriched final states: exclusive single top, top pair and multi-top, or those in association with the SM gauge bosons, leptons and quarks. Motivated by such scenarios, numerous searches have been performed by various experimental collaborations, particularly the CMS and ATLAS. At the LHC, with the exception of a few mild excesses (see \cite{Crivellin:2023zui} for a review), the observations are found to be consistent with the SM expectation. This has led to stringent limits on the new particles' masses and couplings. In fact, in most of the cases, the limits have been pushed to the TeV scale, although not in full generality. For TeV scale states, their decay products---SM leptons, quarks and bosons---could be highly Lorentz-boosted that the jets emanating from them would be collimated. Consequently, the hadronically decaying candidates (primarily, the top quarks in our case) are more likely to manifest as a single fat jet rather than multiple resolved jets. In this section, without pretending to provide an exhaustive and self-sufficient description of the new physics-induced top searches, we briefly discuss various new physics scenarios contributing to the top-enriched final states at the LHC and summarise the relevant LHC searches, including those targeting boosted top quarks in the final state.

\subsection{Extra Gauge Bosons}
\label{sec:ExtraGaugeBosons}
A wealth of BSM models, such as Grand Unification Theories (GUTs) \cite{Georgi:1974sy,Pati:1974yy,Fritzsch:1974nn,Gursey:1975ki,Achiman:1978vg,Ma:1986we,Hewett:1988xc}, supersymmetry \cite{Volkov:1973ix,Wess:1974tw,Wess:1974jb,Ferrara:1974pu,Salam:1974ig,Babu:1987kp}, wrapped extra-dimensions \cite{Randall:1999ee,Cheng:2001vd,Appelquist:2000nn,Burdman:2006gy}, technicolour models \cite{Chivukula:1995gu,Malkawi:1996fs,Hill:2002ap}, little Higgs theories \cite{Weinberg:1972fn,Georgi:1975tz,Georgi:1974yw,Arkani-Hamed:2001nha,Schmaltz:2005ky} and theories featuring dynamical or elaborated spontaneous symmetry breaking \cite{Mohapatra:1974gc,Weinberg:1975gm,Senjanovic:1975rk,Beg:1977ti,Mohapatra:1977mj,Susskind:1978ms,Senjanovic:1978ev,Dimopoulos:1979es,Eichten:1979ah,Babu:1987kp,Ma:2010us,Hsieh:2010zr}, envisage the presence of extra gauge bosons ($W^\prime, Z^\prime$), with properties similar or different to those of the SM gauge bosons ($W, Z$). Specifically, $Z^\prime$ bosons require the SM to be superseded with at least an extra $U(1)$ symmetry, while $W^\prime$ bosons require at least an extra $SU(2)$ gauge group. Some theories assume (or motivate) preferential couplings of $W^\prime, Z^\prime$ bosons to the top-quark (or the third-generation fermions). Examples of such theories include the top-colour \cite{Hill:1991at,Hill:1994hp,Dobrescu:1997nm,Georgi:2000wt} and top-flavour \cite{Malkawi:1996fs,Muller:1996dj,He:1999vp} models, and gauged-flavour symmetry models \cite{Burdman:2000yq}. Some simplified models for dark matter (DM) also predict $Z^\prime$ bosons mediating the interactions between DM and normal matter, see Ref.~\cite{Albert:2017onk} and references therein. Moreover, as possible explanation for the recent flavour anomalies \cite{BaBar:2012obs,Belle:2016dyj}, such models have been pursued with great interest, see for example \cite{Abdullah:2018ets}. 

\begin{figure}[!htb]
	\centering
	\includegraphics[width=0.49\textwidth]{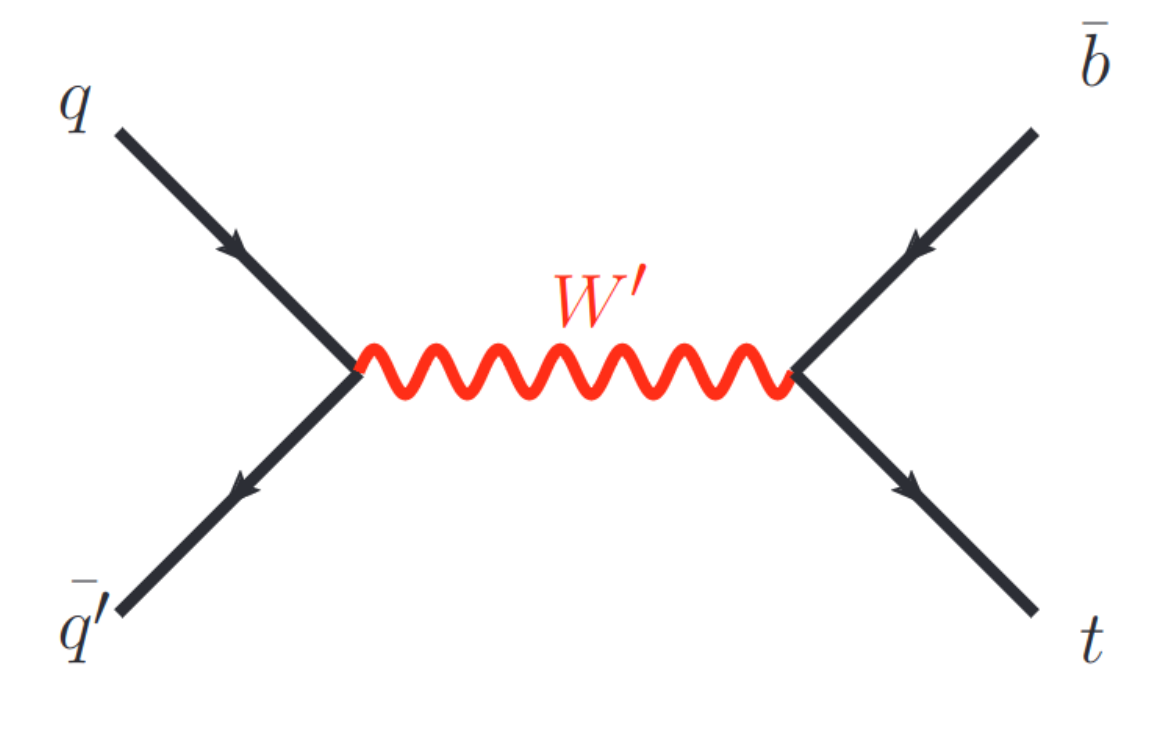}
	\includegraphics[width=0.49\textwidth]{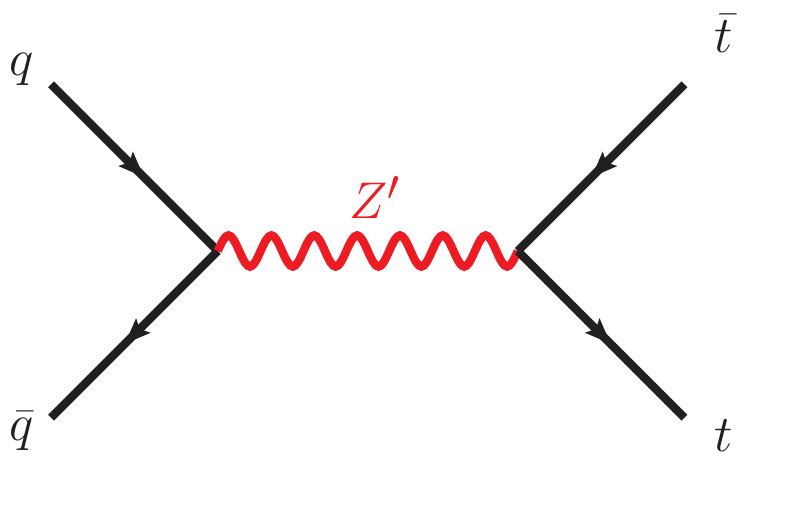}
	\caption{Feynman diagrams for the production of $W^\prime$ \cite{ATLAS:2023ibb} and $Z^\prime$ \cite{ATLAS:2018rvc} bosons and their decays to the third-generation of quarks.}
	\label{fig:WZprime}
\end{figure}

Motivated by such theories, several dedicated searches for $W^\prime$ with right-handed or/and left-handed charged current interactions and $Z^\prime$ bosons decaying to the third-generation quarks (a top-quark and a bottom-quark or a pair of top-quarks, see Figure~\ref{fig:WZprime}) have been performed by both the CMS and ATLAS Collaborations; see for example \cite{CMS:2017zod,ATLAS:2018uca,ATLAS:2018wmg,CMS:2021mux,CMS:2023gte,ATLAS:2023ibb} for $W^\prime$ searches, and \cite{CMS:2017ucf,ATLAS:2018rvc,CMS:2018rkg,ATLAS:2019npw,ATLAS:2020lks,ATLAS:2021mxl,ATLAS:2023taw} for $Z^\prime$ searches. Considering leptonic final states, CMS has excluded left- and right-handed $W^\prime$ bosons with mass below 3.9 and 4.3 TeV, respectively, at the 95\% confidence level (CL)\cite{CMS:2023gte}.\footnote{All the limits quoted here are valid under the assumption that the new particle has a narrow decay width, and they have SM-like couplings.} On the contrary, with the all-hadronic final state considered by CMS, the resulting limits are 3.4 TeV for  both left- and right-handed $W^\prime$ bosons \cite{CMS:2021mux}. Considering both leptonic and all-hadronic final states, ATLAS has provided the most stringent limits of 4.2 and 4.6 TeV for left- and right-handed $W^\prime$ bosons, respectively \cite{ATLAS:2023ibb}. Of the $Z^\prime$ searches targeting the topcolor-assisted-technicolor model \cite{CMS:2017ucf,ATLAS:2018rvc,CMS:2018rkg,ATLAS:2019npw,ATLAS:2020lks,ATLAS:2021mxl,ATLAS:2023taw}, the most stringent limit of 3.9 TeV has been set by ATLAS \cite{ATLAS:2020lks} considering fully hadronic final states. While for the $Z^\prime$ searches targeting the simplified DM models \cite{ATLAS:2018rvc,ATLAS:2019npw}, two benchmarks {\tt A1} and {\tt V1} corresponding to $Z^\prime$ as an axial-vector and vector mediators are used, see \cite{Albert:2017onk} for details. The ATLAS search \cite{ATLAS:2018rvc} in the lepton-plus-jets events has excluded $Z^\prime$ up to 1.2(1.4) TeV for {\tt A1} ({\tt V1}) benchmark models, while the ATLAS search \cite{ATLAS:2019npw} in the fully hadronic final state has excluded $Z^\prime$ in the ranges 0.74--0.97 and 2.0--2.2 TeV (0.80--0.92 and 2.0--2.2 TeV) for {\tt A1} ({\tt V1}) benchmark models. Notably, several of these recent $Z^\prime$ searches employ a Deep Neural Network (DNN) to tag the jets initiated by highly Lorentz-boosted top quarks.

\subsection{Kaluza-Klein excitations}
\label{sec:RSModel}
\begin{figure}[!htb]
	\centering
	\includegraphics[width=0.49\textwidth]{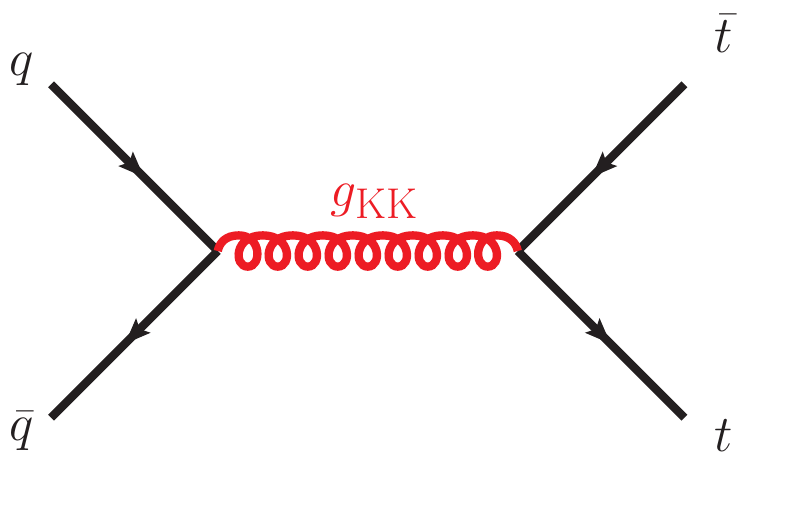}
	\includegraphics[width=0.49\textwidth]{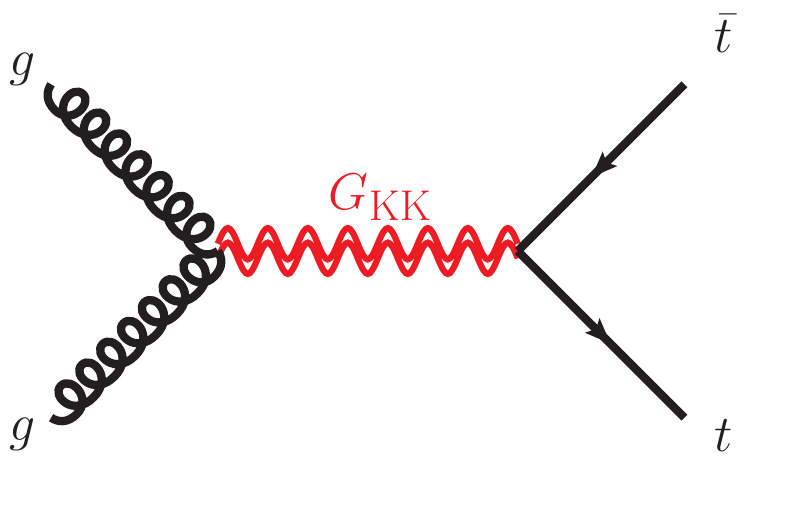}
	\caption{Feynman diagrams for the production of the KK excitations of gluons and gravitons and their decays to the third-generation of quarks \cite{ATLAS:2018rvc}.}
	\label{fig:KK}
\end{figure}

In the Randall-Sundrum (RS) model of wrapped extra dimension, four-dimensional spacetime is embedded in a larger dimensional {\it bulk} with a fifth {\it wrapped} extra dimension \cite{Randall:1999ee}. The propagation of a field in the finite extra dimension manifests as the Kaluza-Klein (KK) excitations in the four-dimensional effective theory. The KK excitations appear as particles with the same quantum number as the original particle but with larger masses. The KK excitations of gluons, gauge bosons and graviton are likely to be localised close to the TeV brane, thereby leading to preferential couplings between these modes and the SM top quark \cite{Lillie:2007ve,Lillie:2007yh,Agashe:2006hk,Agashe:2007zd,Agashe:2007zd,Fitzpatrick:2007qr}. At the LHC, the first KK excitation of gluon $g_{KK}$, spin-1 color-octet boson, will have the largest production rate among the KK states and are therefore expected to be the first observed signal of the RS model \cite{Lillie:2007ve,Lillie:2007yh}. Several searches looking for the $g_{KK}$ have been performed by CMS and ATLAS \cite{CMS:2017ucf,CMS:2018rkg,ATLAS:2018rvc,ATLAS:2019npw}. The $g_{KK}$ is primarily produced quark-antiquark annihilation and decays predominantly into $t\bar t$ (see Fig.~\ref{fig:KK}). On the contrary, the first KK excitation of the graviton $G_{KK}$, spin-2 color-singlet boson, predicted in the bulk RS model \cite{Agashe:2006hk,Agashe:2007zd,Agashe:2007zd,Fitzpatrick:2007qr} is mainly produced in gluon-gluon fusion, and dominantly decays into $t\bar t$ for (sub-)TeV masses (see Fig.~\ref{fig:KK}). Several searches looking for the $G_{KK}$ have been performed by CMS and ATLAS \cite{ATLAS:2018rvc,CMS:2018rkg,ATLAS:2019npw}.

Considering dilepton, single-lepton and fully hadronic final states and employing jet substructure techniques optimised for top quarks with high Lorentz boosts, CMS has excluded $g_{KK}$ excitations with mass below 4.55 TeV \cite{CMS:2018rkg}.\footnote{With the couplings described in Ref.~\cite{Ask:2011zs}, the relative decay width of the $g_{KK}$ resonance lies between 10\% and 30\% depending on its coupling to the top quark.} While the ATLAS search \cite{ATLAS:2018rvc} targetting heavy particles decaying into a $t\bar t$ pair in the lepton-plus-jets events has excluded $G_{KK}$ in the 0.45--0.65 TeV mass range,\footnote{For the mass range 0.4--3 TeV, $G_{KK}$ width varies from 3\% to 6\% when the characteristic dimensionless coupling constant $\kappa/\overline M_{\rm Pl}$ is set to 1 ($\kappa$ is the curvature of the warped extra dimension and $\overline M_{\rm Pl} = M_{\rm Pl}/\sqrt{8\pi}$ is the reduced Planck mass). The branching ratio of $G_{KK}$ to $t\bar t$ increases rapidly from 18\% to 50\% for masses between 400 and 600 GeV, plateauing at 68\% for masses larger than 1 TeV.} the ATLAS search \cite{ATLAS:2019npw} in the fully hadronic final state is not sensitive enough to exclude any mass.

\subsection{Extra Scalars}
\label{sec:ExtraScalars}
\begin{figure}[!htb]
	\centering
	\includegraphics[width=0.32\textwidth]{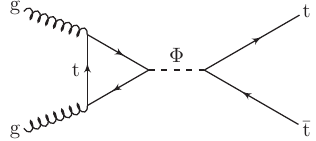}
	\includegraphics[width=0.32\textwidth]{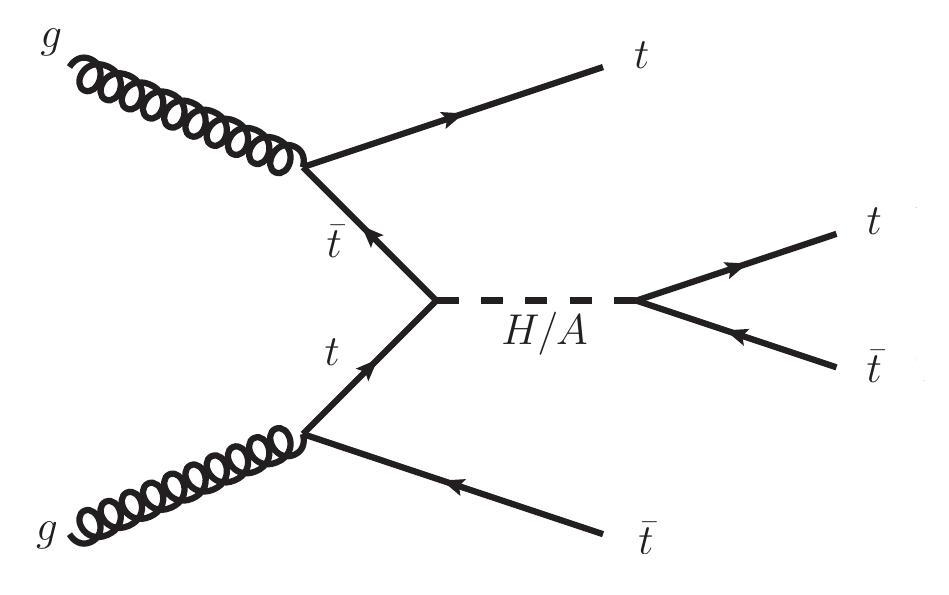}
	\includegraphics[width=0.32\textwidth]{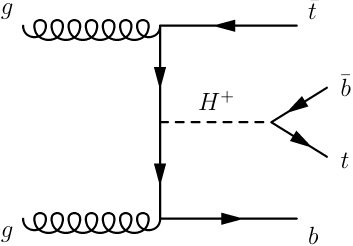}
	\caption{Feynman diagrams for the production of a pseudoscalar or scalar Higgs, denoted by a common symbol $\Phi$ (left) \cite{CMS:2019pzc}, and in association with a pair of top quarks (middle) \cite{CMS:2017tec}, and charged Higgs in association with a top quark and a bottom quark (right) \cite{ATLAS:2018ntn} with the Higgses decaying into third-generation quarks.}
	\label{fig:HA}
\end{figure}

Though the properties of the SM Higgs reported so far have been largely consistent with the SM prediction, the minimality of the SM Higgs sector---the presence of a single $SU(2)_L$ doublet scalar that simultaneously gives mass to the electroweak gauge bosons and all SM fermions---is not guaranteed by any guiding principle or symmetry. As such, a plethora of models with the extended scalar sector have been proposed in the literature, including the addition of $SU(2)_L$ singlets \cite{Silveira:1985rk,Pietroni:1992in,McDonald:1993ex}, doublets \cite{Lee:1973iz,Haber:1984rc,Kim:1986ax,Peccei:1977hh,Turok:1990zg} and triplets \cite{Konetschny:1977bn,Cheng:1980qt,Lazarides:1980nt,Schechter:1980gr,Magg:1980ut,Mohapatra:1980yp}. Many of these models predict new pseudoscalar ($A$) and scalar ($H$) states and sometimes also a charged scalar ($H^\pm$) coupling strongly to the third-generation quarks (see Fig.~\ref{fig:HA}). The widely studied example of such a model is the two-Higgs-doublet models (2HDMs) \cite{Branco:2011iw}. In particular, the type-II variant of 2HDMs, akin to the Higgs sector of the minimal supersymmetric standard model (MSSM) \cite{Fayet:1976et,Fayet:1977yc,Farrar:1978xj,Fayet:1979sa,Dimopoulos:1981zb,Gunion:2002zf,Branco:2011iw,Djouadi:2013uqa}, predict such states predominantly decaying into $t\bar t$ for $m_{A,H} \gtrsim 500$ GeV and small $\tan\beta$ (where $\tan\beta$ is the ratio of the vacuum expectation values of the two Higgs fields).

Both ATLAS \cite{ATLAS:2017snw} and CMS \cite{CMS:2019pzc} have performed searches for $H/A$ decaying into $t\bar t$. Interpreted within the context of a $CP$-conserving type-II 2HDM with a softly broken $Z_2$ symmetry \cite{Gunion:2002zf} in the alignment limit \cite{BhupalDev:2014bir}, the ATLAS search \cite{ATLAS:2017snw} tightens significantly the model parameter space in the high mass ($m_{A,H} \gtrsim 500$ GeV) and low $\tan\beta$ ($\tan\beta$ below 0.69--1.55 depending on $m_{A,H}$) region. Likewise, the CMS search \cite{CMS:2019pzc} excludes the region with $\tan\beta$ below 1.0--1.5 in the 400--700 GeV range for $m_A$ in the {\it habemus} MSSM (hMSSM) model \cite{Djouadi:2013uqa}. Notably, a moderate pseudoscalar Higgs signal-like deviation is observed at $m_A \approx 400$ GeV, with the significance amounting to 1.9$\sigma$. Several searches for $H/A$, produced in association with a pair of top quarks, with $H/A$ decaying into $t\bar t$ (see Fig.~\ref{fig:HA}) have also been performed \cite{CMS:2017tec,CMS:2019rvj,ATLAS:2022rws}. Assuming that either $H$ or $A$ contributes to the four-top final state, the ATLAS search \cite{ATLAS:2022rws} excludes $\tan\beta$ below 1.2--0.5 in the 400--1000 GeV mass range. This exclusion range increases to $\tan\beta$ below 1.6--0.6 when both $H$ and $A$ contribute to the final state. On the other hand, the CMS search \cite{CMS:2019rvj} excludes $H(A)$ in the 350--470 (350--550) GeV mass range for $\tan\beta = 1$ in the alignment limit \cite{BhupalDev:2014bir} of the type-II 2HDM \cite{Branco:2011iw}. Finally, $H^\pm$ heavier than the top quark and decaying into a top quark and a bottom quark (see Fig.~\ref{fig:HA}) are also searched for by CMS and ATLAS \cite{ATLAS:2018ntn,CMS:2020imj,ATLAS:2021upq}. Of these searches, the most recent one by ATLAS \cite{ATLAS:2021upq} excludes a broader region of the $\tan\beta$--$m_{H^\pm}$ plane in the context of the hMSSM and several other scenarios within the MSSM. It excludes $\tan\beta$ in the range 0.5--2.1 for $m_{H^\pm}\in [200-1200]$ GeV . For $m_{H^\pm} \in [200-740]$ GeV, values of $\tan\beta > 34$ are also excluded.

It must be emphasised that none of these searches considers fully hadronic final states, let alone employing jet substructure or machine learning techniques to tag the top-quark initiated jets.

\subsection{Leptoquarks}
\label{sec:LQ}
Similarities between quarks and leptons in the SM, such as their transformations under the SM gauge groups, the number of generations and the hierarchy across generations, motivate a fundamental symmetry connecting them. Such symmetries are embedded in many BSM models, including GUTs \cite{Georgi:1974sy,Pati:1974yy,Fritzsch:1974nn}, technicolour models \cite{Dimopoulos:1979sp}, or theories of quark and lepton compositeness \cite{Buchmuller:1986iq}. Such models predict the existence of ``leptoquarks'' (LQs) with spin-0 (scalar LQs) or spin-1 (vector LQs) that couple to both leptons and quarks simultaneously. LQs transform as triplets under the $SU(3)_C$ and carry fractional electric charges. For vector LQs, in addition, the coupling strength depends on the anomalous magnetic moment ($\kappa$). The $\kappa = 1$ limit refers to the Yang-Mills-type coupling scenario, and the $\kappa = 0$ limit refers to the minimal vector coupling scenario, where the Yang-Mills-type couplings are turned off \cite{Blumlein:1996qp}. While in the minimal Buchmuller-Ruckl-Wyler (BRW) model \cite{Buchmuller:1986zs}, LQs are assumed to couple only to leptons and quarks from the same generation, cross-generational couplings with varying strengths are also possible. As such, LQs can generate lepton flavour universality-violating (LFUV) interactions and, therefore, have been pursued with great interest as a viable solution to the $b$-anomalies \cite{Hiller:2014yaa,Bauer:2015knc,DiLuzio:2017chi,Buttazzo:2017ixm} and the muon's anomalous magnetic dipole moment anomaly \cite{Hiller:2018wbv,Camargo-Molina:2018cwu}.

\begin{figure}[!htb]
	\centering
	\includegraphics[width=0.24\textwidth]{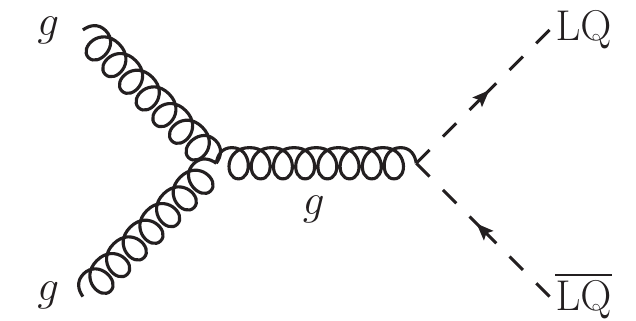}
	\includegraphics[width=0.24\textwidth]{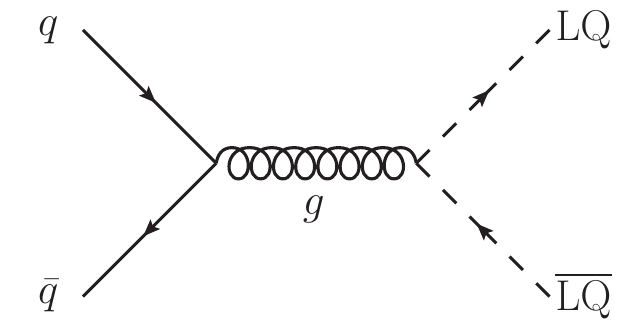}
	\includegraphics[width=0.24\textwidth]{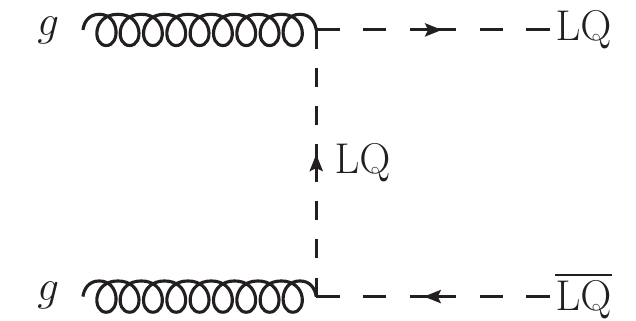}
	\includegraphics[width=0.24\textwidth]{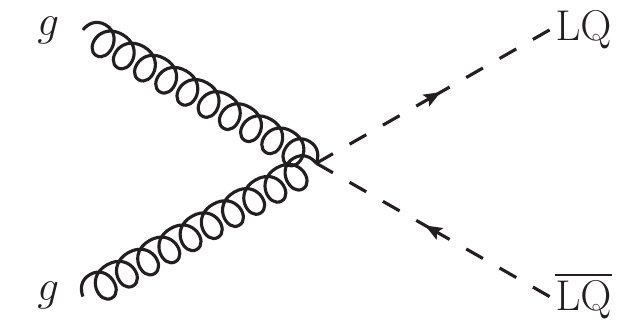}
	\caption{Leading order Feynman diagrams for the production of LQ pairs at the LHC \cite{CMS:2018svy}.}
	\label{fig:LQ}
\end{figure}

At the LHC, LQs are produced singly or in pairs. Pair production proceeds via gluon–gluon fusion and quark-antiquark annihilation mediated by the strong interaction (see Fig.~\ref{fig:LQ}). Several searches have been performed to find LQs decaying to a top quark and a lepton \cite{CMS:2018oaj,CMS:2018svy,ATLAS:2019qpq,CMS:2020wzx,ATLAS:2021oiz,CMS:2022nty,ATLAS:2020dsf,ATLAS:2021jyv}. Among those targeting third-generation down-type scalar LQs decaying into a top quark and a $\tau$-lepton, the ATLAS search \cite{ATLAS:2021oiz} provides the most stringent limit of 1.43 TeV. The upper limit on the mass of third-generation up-type scalar LQ decaying into a top quark and a neutrino is 1.24 TeV \cite{ATLAS:2020dsf}. For cross-generational scalar LQs decaying into a top and an electron (a muon), the limit is 1.58 (1.59) TeV \cite{ATLAS:2023prb}.

The CMS searches \cite{CMS:2020wzx} and \cite{CMS:2018oaj}, respectively, have excluded third-generation up-type and down-type vector LQs below 1.29(1.65) and 1.19(1.47) TeV mass assuming the $\kappa = 0$ ($\kappa = 1$) coupling scenario. For cross-generational vector LQs decaying into a top and an electron or a muon, both ATLAS \cite{ATLAS:2023prb} and CMS \cite{CMS:2018oaj} have put similar limits of approximately 1.7(2.0) TeV in the $\kappa = 0$ ($\kappa = 1$) coupling scenario.

\begin{figure}[!htb]
	\centering
	\includegraphics[width=0.49\textwidth]{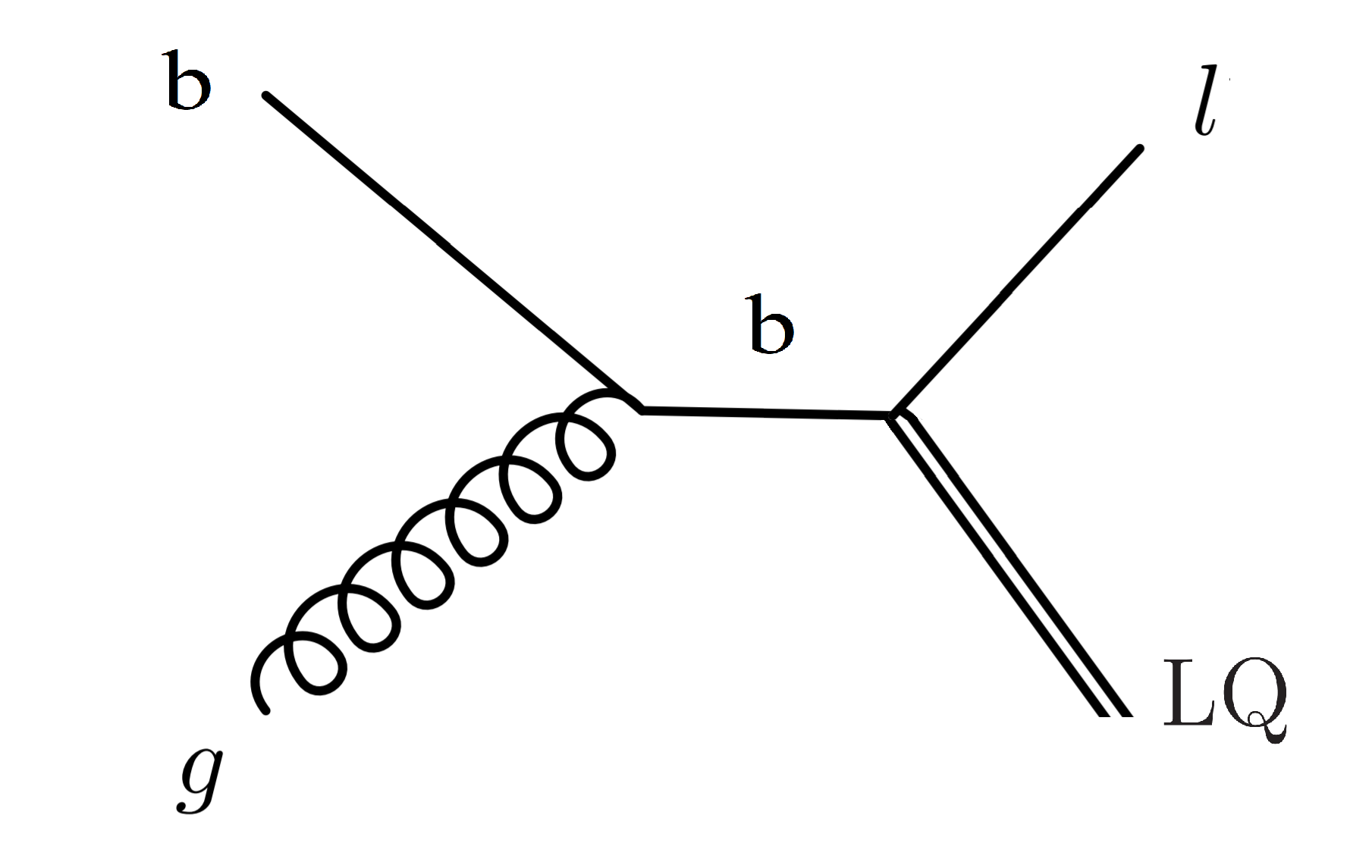}
	\caption{Feynman diagram for the single production of LQs, in association with a lepton, at the LHC \cite{CMS:2020wzx}.}
	\label{fig:LQs}
\end{figure}

Before closing this section, we briefly mention single LQ searches. For single LQ production in association with a lepton (see \ref{fig:LQs}), the cross-section depends on the strength of the quark-lepton-LQ interaction, $\lambda$. Most of the single LQ searches so far have targetted LQs coupling to the $\tau$-lepton and the $b$-quark \cite{CMS:2018txo,ATLAS:2021jyv,CMS:2022ncp,ATLAS:2023kek,CMS:2023bdh}, with the exception of the CMS search \cite{CMS:2020wzx} which assumes equal couplings for the scalar (vector) LQ to $t\tau$ and $b\nu$ ($t\nu$ and $b\tau$). When only the single production of LQs is considered, for $\lambda = 1.5 (2.5)$, this search excludes scalar LQ below 0.55(0.75) TeV and vector LQs below 1.03(1.25) [1.20(1.41)] TeV in the $\kappa = 0$ [$\kappa = 1$] coupling scenario.

Once again, it must be mentioned that none of the experimental searches targeting LQs fully exploit the jet substructure or machine learning techniques to tag the jets initiated by boosted top-quark.

\subsection{Vector-like Quarks}
\label{sec:VLQ}
\begin{figure}[!htb]
	\centering
	\includegraphics[width=0.32\textwidth]{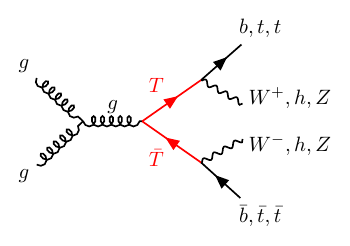}
	\includegraphics[width=0.32\textwidth]{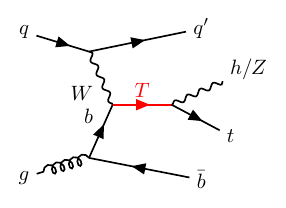}
	\includegraphics[width=0.32\textwidth]{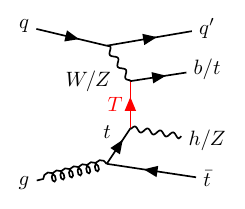}
	\caption{Leading order Feynman diagrams for the pair (left) \cite{ATLAS:2018ziw} and single (middle \cite{ATLAS:2022ozf} and right \cite{ATLAS:2023bfh}) production of VLQs at the LHC.}
	\label{fig:VLQ}
\end{figure}

Attempts to alleviate the {\it fine-tuning} or {\it naturalness} or {\it hierarchy} problem of the SM have led to the idea that the Higgs boson is a composite particle generated by a new strongly interacting sector at the {\it compositeness} scale much larger than the electroweak (EW) scale, and the gap between these two scales is explained by interpreting the Higgs boson as a pseudo-Nambu–Goldstone boson (pNGB) associated with spontaneous symmetry breaking at the {\it compositeness} scale \cite{Dimopoulos:1981xc,Kaplan:1983fs,Kaplan:1983sm,Dugan:1984hq,Perelstein:2005ka,Susskind:1978ms,Hill:2002ap,Kaplan:1991dc}. On this idea, based are the Composite Higgs models \cite{Dimopoulos:1981xc,Kaplan:1983fs,Kaplan:1983sm} and Little Higgs models \cite{Weinberg:1972fn,Georgi:1975tz,Georgi:1974yw,Arkani-Hamed:2001nha}. Such models usually follow a sequential symmetry-breaking pattern: a large global symmetry group above the compositeness scale is spontaneously broken to a smaller group, which is explicitly broken to the SM EW group. Such models, therefore, naturally accommodate several new particles, in particular, new fermionic resonances called vector-like quarks (VLQs): colour-triplet spin-1/2 fermions with both left- and right-handed chiral components transforming identically under $SU(2)$.  Renormalisability and gauge completeness restrict the $SU(2)$ representation of the VLQs to 1, 2 and 3. Like the SM quarks, VLQs carry fractional electric charges. The VLQs with $+2/3$ electric charge ($T$) can, therefore, decay into $tZ,th$ and $bW$; those with $-1/3$ charge ($B$) decay into $bZ,bh$ and $tW$; those with $+5/3$ charge ($X$) into $tW$; and those with $-4/3$ charge ($Y$) into $bW$. The relative couplings for these interactions are determined by the gauge representation of the vector-like quarks. At the LHC, VLQs are produced singly or in pairs. Pair production proceeds via gluon-gluon fusion, and single productions (both resonant and non-resonant) in association with quarks are mediated by a gauge boson (see Fig.~\ref{fig:VLQ}). Note that VLQs also appear in many other BSM scenarios which are well-placed to address various theoretical as well as experimental issues of the SM. Such models, often, introduce additional matter states for the sake of ultraviolet completion, thereby opening up new production and decay channels. The rate for these new channels depend on the details of the UV-completion, for example, their coupling with additional states, or their effective mixings with the SM counterparts, either through direct Yukawa couplings or through a combination of couplings involving yet other new states. In the following, we briefly discusses the LHC searches targeting the VLQs coupling preferentially to the top quark. 

Several searches for pair-produced VLQs decaying to a $W/Z/h$-boson and a $t/b$-quark have been performed by CMS and ATLAS \cite{ATLAS:2017nap,ATLAS:2018ziw,ATLAS:2018uky,CMS:2019eqb,CMS:2020ttz,CMS:2022fck,ATLAS:2022tla,ATLAS:2022hnn,ATLAS:2024gyc}. Considering events with at least one leptonically decaying $Z$-boson and a $b$-tagged jet, the ATLAS search \cite{ATLAS:2022hnn} has put the most stringent limit of 1.6 TeV on $T$ decaying into $tZ$. On the contrary, the CMS search \cite{CMS:2022fck} with single-lepton, same-sign charge dilepton, and multilepton final states has the most stringent limit of 1.56 TeV on $B$ decaying into $tW$. The same search has also put the most stringent limit of 1.50 TeV on $T$ decaying into $th$. Notably, equally competitive limits of 1.46(1.47) TeV has been put on $T$ decaying into $Zt$ ($B/X$ decaying into $tW$) by another ATLAS search \cite{ATLAS:2022tla}. This analysis is based on a final-state signature with high missing transverse momentum, one lepton, and at least four jets, including a $b$-tagged jet. Before going into the single production searches, we briefly mention the searches with fully hadronic final states \cite{ATLAS:2018uky,CMS:2019eqb}. Of these, the CMS search \cite{CMS:2019eqb}, employing a multiclass classification algorithm 'boosted event shape tagger' to tag the jets emanating from $W,Z,h$-bosons or $t$-quarks, has put limits of 1.26(1.37)[1.23] TeV on $T$ decaying into $tZ$ ($T$ decaying into $th$) [$B$ decaying into $tW$]. Note that these limits are quite weaker than those from the leptonic and semileptonic searches.

For single VLQ production in association with quarks, the cross-section depends on the couplings of the VLQ to third-generation quarks $k_{\rm VLQ}$. In the rest of this section, we briefly summarise some of the most recent LHC searches. The CMS search \cite{CMS:2023agg} targeting single $T$ production with $T \to th$ has put an upper limit of 0.16--0.33 on $k_T$ for 600--1200 GeV mass. This search exploits the Higgs boson decay to a pair of photons and, in addition, requires at least one $b$-tagged jet. On the other hand, the ATLAS search \cite{ATLAS:2022ozf} considers fully hadronic final states from $T \to th$ and uses tagging algorithms to identify the jets emanating from the hadronically decaying $h$ and $t$. This search has put an upper limit of 0.35--1.6 on $k_T$ for 1.0--2.3 TeV mass. For VLQs decaying into $tW$, the CMS search \cite{CMS:2018dcw} with one electron or muon and $t/W/b$-tagged jets in the final state provides the most stringent limits. This analysis excludes $B$ ($X$) with left-handed couplings and a relative width of 10\%, 20\% and 30\% below 1490, 1590 and 1660 GeV (920, 1300 and 1450 GeV), respectively. Finally, for single $T$ production with $T$ decaying into $tZ$, both CMS \cite{CMS:2022yxp} and ATLAS \cite{ATLAS:2023bfh} have published their search based on the full Run2 data. While the CMS search \cite{CMS:2022yxp} considers the final state with jets and missing transverse momentum, with the top quark decaying hadronically and the $Z$-boson decaying to neutrinos, the ATLAS search \cite{ATLAS:2023bfh} targets the final state with a pair of leptons from leptonically decaying $Z$-boson and at least one $b$-tagged jet in addition to forward jets. The ATLAS search excluded $k_T$ below 0.22--0.64 for 1000--1975 GeV masses of the singlet $T$ (0.54--0.88 for 1000--1425 GeV masses of the doublet $T$). The CMS search has excluded singlet $T$ below 0.98, 1.1, 1.3, and 1.4 TeV masses for resonance widths 5, 10, 20, and 30\% of the mass, respectively.

\subsection{Supersymmetry}
\label{sec:SUSY}
Supersymmetry (SUSY) theories have been accepted as one of the most motivated theoretical constructs going beyond the SM. For one, SUSY, an extension of the Poincar\'e space-time symmetry, relates bosons and fermions, thereby, as a consequence, solving the hierarchy problem of the SM. SUSY theories are promising candidates for a unified theory, as, within such theories, the measured gauge couplings extrapolated from the EW scale through the SUSY renormalization group equations unify at a GUT scale. Also, the lightest SUSY particle can serve as a DM candidate in the presence of an additional conserved quantum number, such as $R$-parity\footnote{$R$-parity is defined as $R=(-1)^{3(B-L)+2S}$, where $S$ is the spin of the particle, and $B$ and $L$ are the baryon and lepton numbers. The SM particles have $R$-parity of $+1$, while their SUSY partners have $-1$.} that distinguishes SM states from their SUSY partners. Moreover, being the only possible extension of Poincar\'e space-time symmetry group, SUSY is a very likely description of Nature if one considers the sheer elegance of SUSY as a theory.

The SUSY phenomenology, to a large extent, is driven by the presence of $R$-parity. In most of the SUSY models, as we do here, the conservation of $R$-parity is assumed. The $R$-parity-violating scenarios are discussed in a separate section (see Sec.~\ref{sec:RPV}). This has two significant implications for phenomenology. For one, SUSY particles are produced only in pairs. Second, it ensures the stability of the lightest SUSY particle (LSP), thus implying that each SUSY particle produced will entail the LSP at the end of its decay chain. The LSPs, much like the SM neutrinos, leave the detector undetected. Therefore, their presence is usually sought for as missing transverse momentum $p_T^{\rm miss}$. Note that, depending on the underlying mechanism of SUSY breaking, different models predict different SUSY particles as the LSP, thereby leading to different phenomenology. For example, the minimal super-gravity models \cite{Chamseddine:1982jx,Barbieri:1982eh,Ibanez:1982ee,Hall:1983iz,Ohta:1982wn,Kane:1993td} predict the lightest neutralino $\tilde \chi_1^0$ as the LSP, while the gauge-mediated SUSY breaking models \cite{Dine:1981za,Dimopoulos:1981au,Alvarez-Gaume:1981abe,Nappi:1982hm} predict the nearly massless gravitino as the LSP. A detailed discussion of various models is beyond the scope of this paper. The phenomenological mentions hereinafter is based on the simplified models proposed by the LHC New Physics Working Group \cite{LHCNewPhysicsWorkingGroup:2011mji} and the phenomenological minimal supersymmetric model (pMSSM) \cite{Cahill-Rowley:2012ydr}. In particular, the scenarios leading to top-enriched final states are mentioned. And, unless otherwise stated, the lightest neutralino $\tilde \chi_1^0$ is assumed to be the LSP.

\begin{figure}[!htb]
	\centering
	\includegraphics[width=0.24\textwidth]{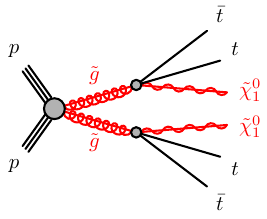}
	\includegraphics[width=0.24\textwidth]{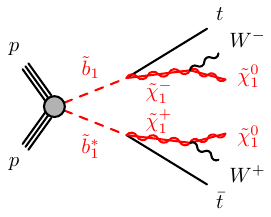}
	\includegraphics[width=0.24\textwidth]{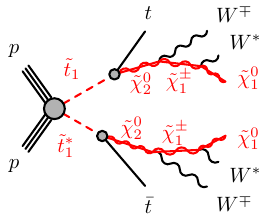}
	\includegraphics[width=0.24\textwidth]{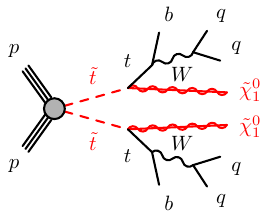}
	\caption{Representative decay topology for pair production of gluino and squarks \cite{ATLAS:2022ihe, ATLAS:2019fag, ATLAS:2020dsf}.}
	\label{fig:SUSY}
\end{figure}

Among the SUSY productions, the pair production of gluinos ($\tilde g$) dominates for a given mass scale, followed by squarks ($\tilde q$), electroweakinos ($\tilde \chi^{0,\pm}$), sleptons ($\tilde \ell^\pm$), and sneutrinos ($\tilde N$). Both CMS and ATLAS have performed numerous dedicated searches looking for the SUSY particles \cite{CMS:2016igb,CMS:2016eep,CMS:2016sth,ATLAS:2016dlg,CMS:2017gbz,CMS:2017jrd,CMS:2017mbm,CMS:2017okm,CMS:2017qxu,ATLAS:2017drc,ATLAS:2017eoo,ATLAS:2017www,ATLAS:2017tmw,ATLAS:2018qzb,CMS:2019ysk,CMS:2019zmd,ATLAS:2019gdh,ATLAS:2019fag,CMS:2019zmd,CMS:2019ybf,ATLAS:2020xgt,ATLAS:2020dsf,ATLAS:2020xzu,CMS:2020cur,CMS:2020cpy,CMS:2021beq,ATLAS:2021pzz,ATLAS:2021yij,ATLAS:2021kxv,ATLAS:2021hza,ATLAS:2023dbq}. Of these, the ATLAS search \cite{ATLAS:2022ihe} targeting gluinos decaying into a pair of top-quarks and the lightest neutralino $\chi_1^0$ (see Fig.~\ref{fig:SUSY}) in final states with missing transverse momentum and three or more $b$-jets, has put the most stringent limit of 2.44 TeV. On the contrary, among the searches targeting third-generation squarks $\tilde b_1$ and $\tilde t_1$, the ATLAS searches \cite{ATLAS:2019fag} and \cite{ATLAS:2020dsf} are the most constraining ones. The search in Ref.~\cite{ATLAS:2019fag}, considering final states with same-sign leptons and energetic jets, excludes $\tilde b_1$ decaying into a top quark, a $W$ bosons and the LSP $\tilde \chi_1^0$ (see the second diagram in Fig.~\ref{fig:SUSY}) up to 750 GeV. The same search also excludes $\tilde t_1$ decaying into a top quark, a $W$ bosons and $\tilde \chi_1^\pm$, with $\tilde \chi_1^\pm$ further decaying into the LSP $\tilde \chi_1^0$ and an off-shell $W$-bosons (see the third diagram in Fig.~\ref{fig:SUSY}) up to 750 GeV; while the search in Ref.~\cite{ATLAS:2020dsf} has excluded $\tilde t_1$ decaying into a top quark and the LSP $\tilde \chi_1^0$ up to 1.25 TeV, considering all hadronic $t\bar t$ plus missing transverse momentum final state. Notably, this search exploits fatjet properties to efficiently reconstruct top quarks that are Lorentz-boosted in the laboratory frame.

\subsection{$R$-parity-violating Supersymmetry}
\label{sec:RPV}

The previous section extensively discussed the existing collider constraints on $R$-parity-conserving SUSY scenarios. The primary motivation behind introducing $R$-parity was to prevent rapid proton decay. This ad hoc symmetry makes the lightest supersymmetric particle (LSP) stable, potentially qualifying it as a dark matter candidate \cite{PhysRevLett.50.1419,Ellis:1983ew}. However, the rationale behind enforcing $R$-parity lacks a theoretical foundation, and SUSY theories contravening this symmetry are equally compelling. Supersymmetric theories violating $R$-parity exhibit more natural mass spectra and encounter fewer experimental constraints than their $R$-parity-conserving counterparts. This fact has motivated the ATLAS \cite{Aaboud:2017faq, Aad:2020uwr,Aad:2019ftg,Aaboud:2018zeb,Aaboud:2018lpl,SUSY-2016-09,Aaboud:2017opj,SUSY-2016-14,ATLAS:2017jnp,ATLAS:2017jvy,ATLAS:2017oes} and CMS collaboration \cite{Sirunyan:2017dhe,Khachatryan:2016iqn,Chatrchyan:2013xsw,CMS:2021knz,CMS:2018mts,CMS:2014wpz,CMS:2016ooq} to conduct several searches of RPV SUSY scenarios. In the following, we briefly list a few such searches. \\

\begin{figure}[!htb]
	\centering
	\begin{subfigure}[t]{0.32\linewidth}
		\includegraphics[width=1\textwidth]{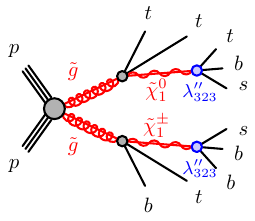}
		\caption{}
		\label{fig:Gtt}
	\end{subfigure}
	\begin{subfigure}[t]{0.32\linewidth}
		\includegraphics[width=1\textwidth]{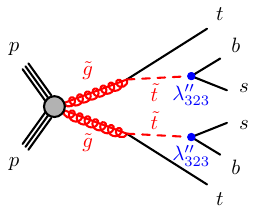}
		\caption{}\label{fig:Gtbs}
	\end{subfigure}
	\begin{subfigure}[t]{0.32\linewidth}
		\includegraphics[width=1\textwidth]{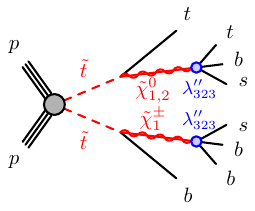}
		\caption{}\label{fig:stop}
	\end{subfigure}
	\begin{subfigure}[t]{0.32\linewidth}
		\includegraphics[width=1\textwidth]{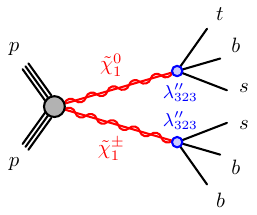}
		\caption{}\label{fig:C1N1}
	\end{subfigure}
	\begin{subfigure}[t]{0.32\linewidth}
		\includegraphics[width=1\textwidth]{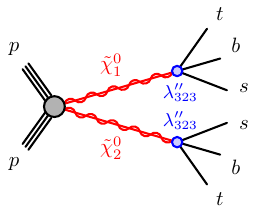}
		\caption{}\label{fig:N1N2}
	\end{subfigure}
	\begin{subfigure}[t]{0.32\linewidth}
		\includegraphics[width=1\textwidth]{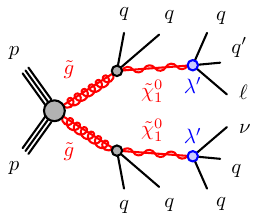}
		\caption{}\label{fig:LQD}
	\end{subfigure}
	\caption{Examples of signal diagrams for the simplified RPV models considered in the ATLAS analysis \cite{ATLAS:2021fbt}. For simplicity, particles and anti-particles are shown using the same symbols, omitting the anti-particle notation.}
	\label{fig:diagram_A1}
\end{figure}

Looking into the final state with a single isolated lepton (either an electron or a muon) accompanied by numerous jets, some of which may be b-tagged, the ATLAS collaboration \cite{ATLAS:2021fbt} has effectively constrained various simplified RPV SUSY scenarios, as illustrated in Figure \ref{fig:diagram_A1}. Within models where $\Tilde{g}\rightarrow t \Bar{t}\Tilde{\chi}_1^0\rightarrow t \Bar{t} t b s$ (depicted in Fig. \ref{fig:Gtt}), gluino masses of up to 2.38 TeV have been excluded at a confidence level of 95\%. Similarly, for scenarios involving direct stop production (as depicted in Fig. \ref{fig:stop}), stop masses up to 1.36 TeV have been ruled out at a 95\% confidence level. In models where $\Tilde{g} \rightarrow \Bar{t}\Tilde{t}$ and $\Tilde{t} \rightarrow \Bar{b}\Bar{s}$ (illustrated in Fig. \ref{fig:Gtbs}), as well as $\Tilde{g}\rightarrow q \Bar{q}\Tilde{\chi}_1^0\rightarrow q \Bar{q} q \Bar{q} l/\nu$ (shown in Fig. \ref{fig:LQD}), gluino masses up to 1.83 TeV and 2.25 TeV, respectively, have been excluded at a confidence level of 95\%. Additionally, for models involving direct electroweakinos production (as portrayed in Fig. \ref{fig:C1N1} and \ref{fig:N1N2}), Higgsino (Wino) masses ranging from 200 (197) GeV to 320 (365) GeV have been excluded. Pair production of stops with subsequent decay into the top quark and the lightest neutralino (which further decays into three light quarks) is also considered by the CMS analysis \cite{CMS:2021knz}.  This search has led to the exclusion of top squark masses up to 670 GeV at a confidence level of 95\%.\\

\begin{figure}[htbp]
	\centering
	\subfloat[]{
		\includegraphics[width=0.45\textwidth]{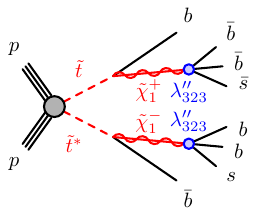}
	}
	\subfloat[]{
		\includegraphics[width= 0.45\textwidth]{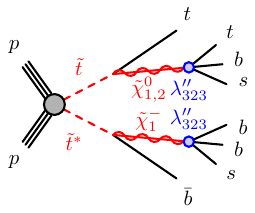}
	}
	\caption{ Diagrams of the signal processes involving pair production of top squarks $\tilde{t}$ \cite{ATLAS:2020wgq}.}
	\label{fig:diagram_A2}
\end{figure}

The ATLAS analysis \cite{ATLAS:2020wgq} has considered scenarios with stops as the lightest colored SUSY particles where the LSP is assumed to be a triplet of two neutralinos ($\tilde{\chi}_1^0,\tilde{\chi}_2^0$) and one chargino ($\tilde{\chi}_1^{\pm}$) states that are mass-degenerate and carry dominantly higgsino components. Here, the strong production of stop pairs can give rise to a final state with a high jet multiplicity (See Figure \ref{fig:diagram_A2}). The absence of any significant deviations from the Standard Model predictions prompts the establishment of a 95\% confidence level upper limit on the stop mass, capped at 950 GeV within the region where $m_{\tilde{t}} - m_{\tilde{\chi}_{1,2}^0,\Tilde{\chi}_1^{\pm}} \le m_{\rm top}$, an exclusive sensitivity zone for this analysis. Meanwhile, the CMS collaboration also investigates stop pair production \cite{CMS:2018pdq}, where each stop decays into four quarks via an intermediate Higgsino-like LSP. This analysis has effectively ruled out stop masses ranging from 100 to 720 GeV, with the Higgsino mass set to 75\% of the stop mass.\\

\begin{figure}[!htb]
	\centering
	\begin{subfigure}[t]{0.42\linewidth}
		\includegraphics[width=1\textwidth]{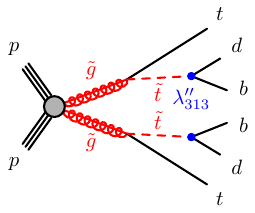}
		\caption{}
		\label{fig:A31}
	\end{subfigure}
	\begin{subfigure}[t]{0.42\linewidth}
		\includegraphics[width=1\textwidth]{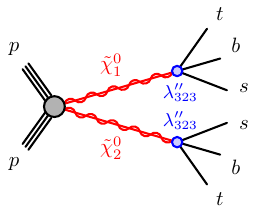}
		\caption{}\label{fig:A32}
	\end{subfigure}
	\caption{Examples of signal diagrams for the simplified RPV models considered in the ATLAS analysis \cite{ATLAS:2021fbt}. For simplicity, particles and anti-particles are shown using the same symbols, omitting the anti-particle notation.}
	\label{fig:diagram_A3}
\end{figure}

The ATLAS analysis \cite{ATLAS:2019fag} explored a final state characterized by same-sign leptons, multiple jets, and significant missing transverse momentum (depicted in Figure \ref{fig:A31}). This search excludes gluino masses below 1.6 TeV for $\tilde{t}$ masses up to 1.2 TeV. Subsequently, another ATLAS study \cite{ATLAS:2023afl} revisited a similar final state, further refining the analysis. Here, the previously established bound on gluino mass is extended to 1.65 TeV, with a Stop mass requirement below 1.45 TeV. The ATLAS analysis \cite{ATLAS:2023lfr} has also looked into the same sign lepton final state but for the case of a higgsino-like $\Tilde{\chi}_1^0/\Tilde{\chi}_2^0$ (See Figure \ref{fig:A32}). The search excludes $\Tilde{\chi}_1^0/\Tilde{\chi}_2^0$ masses up to 200 GeV. Furthermore, the CMS analysis \cite{CMS:2020cpy} also investigated the same-sign lepton final state within two simplified RPV scenarios. Both scenarios entail strong production of gluino pairs. In the first scenario, each gluino is assumed to decay into four quarks and a lepton. The non-observation of any excess over the SM background allowed the analysis to exclude gluino mass up to 2.1 TeV. Conversely, the second scenario involves gluino decay in the $\Tilde{g}\rightarrow t b s$ channel. This scenario was able to rule out gluino mass up to 1.7 TeV.

\section{Summary and Outlook}

The application of advanced machine learning techniques in the field of top tagging has seen significant progress in recent years. In this review, we have tried to summarise some of the recent developments in top tagging algorithms and their possible applications in the field of high-energy physics.

The paper begins with a discussion of various top taggers, focusing on high-level feature-based classifiers, Convolutional Neural Networks (CNNs), and Graph Neural Networks (GNNs). We have hand-picked some algorithms in each category and demonstrated their performance for top tagging. It's important to note that all findings discussed herein are borrowed from their respective sources. Due to variations in datasets used by different algorithms, a direct comparison of their performance isn't feasible. Nevertheless, our exploration offers valuable insights into their efficacy for the task at hand.

The second part of the paper discusses various BSM scenarios that can lead to a final state topology with boosted top quarks at the LHC. We have also discussed the bounds on these scenarios from previous collider searches. In all these cases, an efficient Identification of the final state top quark can help drastically reduce the SM background. It can lead to a possible discovery or even stronger constraints on the model parameter space. We must mention that numerous studies have already implemented different top tagging techniques for the study of BSM physics. However, a review of those studies is beyond the scope of our present discussion.

\acknowledgments
The work of S.A.~is partially supported by the National Natural Science Foundation of China under grant No.~11835013.

\section*{Data Availability Statement}
All the results presented in this review are borrowed from their original analysis. Proper citations have been accorded, and we request the interested reader to consult those references for the datasets. For convenience, we have also discussed the prescription of generating the datasets whenever necessary. For studies where public datasets are used by the original work, we have provided the details of the dataset source within the text.

\bibliographystyle{JHEP}
\bibliography{v0}
 
\end{document}